\documentclass[11pt,a4paper]{article}
\usepackage{jheppub} 
\usepackage{graphicx}
\usepackage{bm}

\usepackage{caption}
\usepackage{framed}
\usepackage{xcolor}
\usepackage{rotating}
\usepackage{empheq}
\usepackage{afterpage}

\usepackage{subcaption}

\usepackage[utf8]{inputenc}
\title{\boldmath Recursive Generation of T\comment{edit after finising}he Semi-Classical Expansion in Arbitrary Dimension}

\author{Cihan Pazarba{\c{s}}\i}
\affiliation{Physics Department, Bo\u{g}azi\c{c}i University\\
	34342 Bebek / Istanbul, Turkey}

\emailAdd{cihan.pazarbasi@gmail.com}

\newcommand{\comment}[1]{[$\clubsuit${\ {\bf #1}}]}

\newcommand{\lbar}{\lower0.2ex\hbox{$\mathchar'26$}\mkern-10mu \lambda}

\usepackage{titling}

\def\Im{\mathrm{Im}}
\def\Re{\mathrm{Re}}

\def\mrmd{{\mathrm{d}}}

\def\th{{\mathrm{th}}}
\def\st{{\mathrm{st}}}
\def\nd{{\mathrm{nd}}}
\def\exp{\mathrm{exp}}

\def\a{\alpha}

\def\D{\Delta}

\def\ve{\varepsilon}
\def\f{\phi}
\def\F{\Phi}

\def\g{\gamma}
\def\G{\Gamma}

\def\h{\eta}
\def\k{\kappa}

\def\m{\mu}
\def\n{\nu}
\def\o{\omega}

\def\s{\sigma}
\def\S{\Sigma}
\def\t{\theta}

\def\tU{\tilde{U}}

\def\tD{\tilde{D}}

\def\mbfA{{\mathbf{A}}}

\def\mbfE{{\mathbf{E}}}
\def\mbff{{\mathbf{f}}}
\def\mbfp{{\mathbf{p}}}
\def\mbfx{{\mathbf{x}}}

\def\mbfH{\mathbf{H}}
\def\mbfh{\mathbf{h}}
\def\mbfpi{\bm{\pi}}
\def\bmh{\bm{\h}}
\def\bmF{\bm{\F}}

\def\st{{\mathrm{st}}}  
\def\nd{{\mathrm{nd}}} 
\def\rd{{\mathrm{rd}}}

\def\mcalA{\mathcal{A}}
\def\mcalB{{\mathcal{B}}}

\def\mcalD{{\mathcal{D}}}
\def\mcalE{\mathcal{E}}
\def\mcalH{\mathcal{H}}

\def\mcalP{\mathcal{P}}
\def\mcalS{\mathcal{S}}
\def\mcalV{{\mathcal{V}}}
\def\E{\mathcal{E}}
\def\B{\mathcal{B}}

\def\mcalJ{\mathcal{J}}
\def\mcalK{\mathcal{K}}

\def\mrmD{{\mathrm{D}}}

\def\mbbR{\mathbb{R}}
\def\mbbN{\mathbb{N}}
\def\mbbC{\mathbb{C}}

\def\mbbE{\mathbb{E}}
\def\mbbK{\mathbb{K}}

\def\msfb{\mathsf{b}}

\def\*{\star}

\def\tC{\tilde C}

\def\Tr{{\mathrm{Tr}}}

\def\diag{\mathrm{diag}}

\def\mrmE{\mathrm{E}}
\def\mrmB{\mathrm{B}}

\def\mrmI{\mathrm{I}}
\def\mrmR{\mathrm{R}}

\def\<{\langle}
\def\>{\rangle} 
\def\dee{\partial}

\def\Im{\mathrm{Im}}
\def\Re{\mathrm{Re}}

\usepackage{slashed}

\definecolor{darkbrown}{rgb}{0.4, 0.26, 0.13}
\definecolor{paleblue}{rgb}{0.69, 0.93, 0.93}
\definecolor{lightskyblue}{rgb}{0.53, 0.81, 0.98}
\definecolor{skyblue}{rgb}{0.53, 0.81, 0.92}
\definecolor{darkred}{rgb}{0.55, 0.0, 0.0}
\definecolor{darkblue}{rgb}{0.0, 0.0, 0.55}
\definecolor{darkpastelgreen}{rgb}{0.01, 0.75, 0.24}
\definecolor{lightgreen}{rgb}{0.56, 0.93, 0.56}

\begin{document}

\begin{titlepage}
	\begin{center}{\Large \bf \boldmath Pair Production in Real Proper Time and Unitarity without Borel Ambiguity }
 	\end{center}
	\hrule
	\vskip 1cm
	\centerline{ {{\bf Cihan Pazarbaşı}}}

	\begin{center}
		\textit{Okinawa Institute of Science and Technology Graduate University,\\
			1919-1 Tancha, Onna-son, Kunigami-gun, Okinawa, Japan 904-0495}
			\vskip 0.2cm
		
		\vskip 0.5cm 
		\texttt{cihan.pazarbasi@gmail.com}\\

	\end{center}
	\vskip 1.0cm
	
	\centerline{\bf Abstract}
	\vskip 0.2cm \noindent 
	The pair production of scalar particles in electromagnetic background fields is analyzed using real proper time formulation of 1-loop effective action. After explaining how real proper time formulation keeps unitarity of the particle creation process in an unambiguous way and discussing the (lack of) pair production in uniform (magnetic) electric backgrounds, we apply these ideas to general electric field backgrounds. Our approach is based on a recursive perturbative expansion of the proper time propagator $U(t)$ and we show how the pair production probabilities can be obtained from its singularities which arises upon a Pade summation of the expansion. We computed the pair production probabilities for general space or time dependent electric fields in locally constant approximation and showed they match with the well-known worldline instanton calculations of pulse and periodic backgrounds. Later, for one dimensional periodic electric and  magnetic  backgrounds, we showed how the WKB integrals appear in the classical limit. We linked both cases to specific WKB cycles in different spectral regions of the WKB problem and computed the WKB actions by evaluating the WKB integrals as well as by taking phase space integrals directly along Lefschetz thimbles. Finally, we explain the equivalence of our construction with exact WKB method in the classical limit by showing how the WKB cycles precisely contribute to the pair production process which is in complete agreement with our time dependent setting. 
	
\end{titlepage}
\hrule
\tableofcontents
\vskip 0.5 cm
\hrule
\vskip 1.5 cm

\section{Introduction}

Particle pair production is one of the fundamental predictions of relativistic quantum theories. Early on, it was noticed that in the presence of constant electromagnetic fields, this can be explained by tunneling through a barrier \cite{1931ZPhy...69..742S} as well as by quantum effects added to the classical Lagrangian  \cite{1936ZPhy...98..714H, 2006physics...5038H,Weisskopf:1936hya,Miller:1994bs}. Later, Schwinger \cite{Schwinger:1951nm} systematically formalized the instability of vacuum and the particle pair production in connection with the imaginary parts of the effective Lagrangian in the basis of gauge theories, i.e. QED in his particular example. Since then, the particle production in background electromagnetic fields has been investigated thoroughly for different types of background potentials using different methods. (See \cite{dunne2005heisenberg, Gelis:2015kya,Fedotov:2022ely} for review on the subject.)

From a mathematical perspective, a consequence of the vacuum instability, so that the particle creation, presents itself in the perturbation theory \cite{PhysRev.85.631}: In QED, an expansion in the fine structure constant $\a$ should be divergent, as the theory is ill-defined for negative values of $\a$. Although at first sight, a divergent series seems pathological, now it is understood in the context of resurgence theory that such series encodes all the physical information in a consistent way. (See  \cite{Aniceto:2018bis} for a comprehensive introductory discussion on the subject.)

A common way to decode the information beyond the perturbative expansion is the Borel summation procedure. For exactly solvable QED background fields, this approach was used at 1 loop \cite{Dunne:1999uy,Dunne:2022esi} and 2 loops \cite{Dunne:2002qg,Dunne:2021acr} orders to obtain the imaginary contribution to the effective action, so that the pair production probability. Standard Borel summation method, however, possesses another pathology: The imaginary contribution is multi-valued. This general feature of the Borel procedure stems from the freedom of choice in the analytical continuation directions in the presence of singularities. While these singularities are sources of non-trivial information, which can not be probed by perturbative expansions, the freedom of choice leads to multi-valued physical quantities.




This pathology has a resolution for problems with a stable vacuum where the imaginary part should not exist in the full solution. 
Based on the works of Bogomolny and Zinn-Justin in \cite{Bogomolny:1980ur,Zinn-Justin:1981qzi}, it has been precisely shown in various settings \cite{Aniceto:2013fka,Dunne_2014,Cherman:2014ofa,Kozcaz:2016wvy,Dunne:2016jsr} that the ambiguous imaginary contribution is canceled by inserting additional non-perturbative information. In this way, the resulting quantity becomes real and, maybe more importantly, unique. 

In this paper, we will concentrate on pair production problems where vacua are not always stable. Therefore, one can not rely on the cancellation mechanism of Bogomolny and Zinn-Justin since the imaginary contributions should survive whenever there is vacuum instability. In addition to that, multi-valued pair production probability is not acceptable as one of these results violate the unitarity of the theory. To see how this violation presents itself in the effective action, let us look  at in-out amplitude, i.e. $\mcalA = \<0_+|0_-\>$, 
where $|0_\pm\>$ are the vacua at infinite past and future. In general, the amplitude is a pure phase, 
i.e.,
\begin{equation}\label{Aplitude_EffectiveAction}
	\mcalA= e^{i \G},
\end{equation}
where $\G$ is the effective action of the theory. As long as $\G\in\mbbR$, the transition probability $|\mcalA|^2=1$ and the vacuum is a stable one, i.e. there is no creation of particles. However, when $\G \in \mbbC$, the transition probability becomes
\begin{equation}\label{VacuumInstability}
	|\mcalA|^2 = e^{-2 \Im \G}.
\end{equation} 
Obviously, whenever $\Im \G\neq 0$, the vacuum is not stable and there is a probability of particle creation which is defined as
\begin{equation}\label{CreationProbability}
	\mcalP \sim 1 - 2 \Im \G \leq 1. 
\end{equation}
Note that since the  theory is unitary, $\Im \G$ should not be negative. At first, this seems to be a trivial statement. However, for consistency and completeness, the $\Im \G>0$ condition should be implicit in methods we use, which should also cover the stable cases on an equal footing. 

To our knowledge, in the context of the pair production problem, the unitarity problem was first noticed in \cite{Chadha:1977my} and it was shown using the exact uniform backgroud field solution that the consistent treatment can be achieved with the Schwinger proper time integral but only when the proper time is chosen to be real. This approach was motivated from two observations:
\begin{itemize}
	\item Schwinger's proper time integral at 1 loop order\footnote{ This fact is not limited to the 1-loop order. See \cite{Dunne:2021acr} for an example at 2 loops.} and the Borel integral have the same form.
	\item When the proper time is chosen real, possible analytical continuation directions of the integral are defined by construction.
\end{itemize}

Let us elaborate on these observations: In general, the divergent series for the perturbative effective action is found first, then the Borel procedure is applied to tame the divergence and probe the singularity which leads to the non-perturbative information, i.e the pair production probability in our case. Instead of following this standard path, the first point suggests that it is possible to probe these singularities before taking the proper time integral. While the two approaches are equivalent, in this way, the crucial analytical continuation information would be kept.

The predefined analytical continuation, on the other hand, stems from  utilization of the $i\ve$ prescription. In addition to \cite{Chadha:1977my}, its relation with the resolution of the  Borel ambiguity was also discussed in \cite{Olesen:1977ih,Pazarbasi:2019web}. In order to summarize the main idea briefly, let us consider the Green's operator $G(m^2) = (m^2 + H)^{-1}$ of a relativistic quantum theory which is the derivative of the effective action, i.e. $G(m^2) = \frac{\mrmd \G(m^2)}{\mrmd m^2}$. It is not well defined on the real $m^2$ line \cite{Taylor:72}. Instead, it is defined by an analytical continuation on the upper or lower half of the complex plane by redefining $G^\pm(m^2) = \lim_{\ve \rightarrow 0} (m^2 \pm i\ve +H)^{-1}$. Then, any observable computed using these redefined propagators have a certain predefined analytical continuation which prevents any ambiguity in the limit $\ve\rightarrow 0$. As we will discuss, in proper time formalism, the analytical continuation information can also be also transferred the proper-time integral contour of the 1 loop effective action when the proper time is taken real instead of imaginary. 

One of the main motivation of this paper is applying these ideas to general electric backgrounds with arbitrary dimensional inhomogineities. For this purpose, we investigated the pair-production of scalar particles in general electromagnetic field backgrounds and compute the leading order corrections to the uniform case in locally constant field approximation (LCFA).

For this purpose, we investigated the pair-production of scalar particles in general electromagnetic field backgrounds and compute the leading order corrections to the uniform case in locally constant field approximation (LCFA).
To a great extent, our results matched with \cite{Gusynin:1995bc,Gusynin:1998bt} where the authors generalized Schwinger's approach to non-uniform electromagnetic fields. However, in the case of spatially inhomogeneous electric fields, we found a sign difference. This is a crucial aspect of the space dependent electric fields since it is related to a critical field strength which states a lower bound for the creation of real particles. 
In addition to that, our approach is conceptually different as we only relied on the perturbative expansion of the proper time propagator $U^\pm(t) = e^{\mp i H t} $, i.e. Fourier transformation of $G^\pm(m^2)$ and the Pade summation of the resulting series. In this way, we obtained the non-perturbative pair production rates of general background fields which arises from a singularity of $U^\pm(t)$ located at complex $t\neq 0$ point without any ambiguity. 

Similar approaches were used in \cite{Dunne:1999uy,Dunne:2022esi} where the authors examines the single-pulse electric fields of form $E(x\o) = \E \sinh(\o x)$ and relied on the existence of an exact solution in this case. Moreover, since their focus was on the divergent expansion of the effective action, the imaginary contribution to the effective action upon a summation process suffers from the Borel ambiguity. Our approach offers a way to generalize their procedures to more general electric backgrounds while keeping the unitarity of the process without any ambiguity. In order to get the perturbative expansions, we used a perturbative scheme developed in \cite{Pazarbasi:2020god} for non-relativistic quantum mechanics and adapted it to arbitrary background gauge potentials. Note that this method is analogous to the small time expansion of heat kernels \cite{DeWitt:1975ys,Vassilevich:2003xt,avramidi2019heat} and string inspired methods \cite{Fliegner:1993wh,Fliegner:1997rk,Schubert:2001he,Edwards:2019eby}, but it organizes the expansion via a recursion relation which enables us to compute high enough orders to extract non-perturbative information out of it.

In addition to obtaining unitary results, another objective of the paper is investigating how this perturbative approach is connected to two widely used semi-classical approaches, i.e. Euclidean worldline instanton (WLI) and the WKB method. A WLI appears as a periodic solution to the classical equation motion in path integral formalism. In the presence of electric field backgrounds, they are widely used in computations of the pair production probabilities, which presents itself as the imaginary part of the effective action \cite{ Affleck:1981bma, Dunne:2005sx, Dunne:2006st,Dunne:2006ur,Dumlu:2011cc, Schneider:2014mla, Dumlu:2015paa, Ilderton:2015qda, Akal:2017sbs}. It was shown \cite{Dietrich:2007vw,Dunne:2008zza} that the WLI method is equivalent to the semi-classical construction of Gutzwiller \cite{Gutzwiller:1971fy,Muratore-Ginanneschi:2002sjs}, which is also linked to Lefschetz thimbles \cite{Sueishi:2020rug}. The thimble perspective was shown to be an consistent way to define and compute real time path integrals in various contexts \cite{Tanizaki:2014xba,Feldbrugge:2017kzv,Feldbrugge:2017fcc} including the pair production problems \cite{Rajeev:2021zae}.

The WKB method, on the other hand, is a connection problem based on the Schrödinger type equations and non-trivial information is encoded in integrals over closed paths which we call \textit{WKB cycles} (or cycles in short). It was shown in various settings \cite{Basar_2015,Codesido:2017dns,Codesido:2017jwp,Basar_2017,Fischbach:2018yiu,Raman:2020sgw} that different WKB cycles of a quantum system are not independent of each other which is a sign of underlying resurgent structure of the corresponding theory and closely related to the Bogomolny Zinn-justin mechanism. 

The application of WKB formalism to  pair production problems \cite{Keldysh:1965ojf,Brezin:1970xf,Nikishov:1970br,1972JETP...34..709P,Kim:2000un,Kim:2003qp,Kim:2007pm,Dumlu:2010ua,Taya:2020dco}, on the other hand, is based on the Klein-Gordon equation for the mode functions. In its essence, non-trivial information arises from the Stokes phenomenon and the exponential parts of the pair production probabilities are linked to the integrals over WKB cycles. This is often mentioned as tunneling description of the pair production problems. An equivalence between the actions of the worldline instantons and the WKB integrals for any background field with one dimensional inhomogeneities was argued in \cite{Kim:2019yts}. (See also \cite{Basar_2015} for an example in the presence of periodic electric field backgrounds.) This equivalence was also more precisely discussed in relation with the exact WKB (EWKB) method in \cite{Taya:2020dco} where the authors made a connection between specific WKB cycles in scattering region and WLI. 

In this line of the paper, we first observed that the pair production probability arising from the analytical structure of $U^\pm(t)$ matches with the WLI calculations in \cite{Dunne:2005sx, Dunne:2006st}. This shows that the location of the singularity associated to the pair production process equal to the values of the Euclidean WLI action and verifies an old conjecture by Andr\'e Voros \cite{voros94} 

Later, focusing on time dependent periodic background fields for both electric and magnetic backgrounds, we showed how the leading order WKB integrals arises from our real time perturbative decomposition in $\hbar\rightarrow 0$ limit. In this way, we identified magnetic and electric backgrounds with the specific WKB cycles of below and above barrier top regions of the periodic potential, respectively. We showed that the WKB action corresponding to the electric background can be linked to both of the WKB cycles above the barrier top while the magnetic case is linked to the classically allowed region under the barrier top. Then, we computed the WKB actions by two approaches: $i)$ Taking the WKB integrals along corresponding cycles. $ii)$ Taking phase-space integrals appear in $\hbar\rightarrow 0$ limit of $U^\pm$ along Lefschetz thimbles, which presents a new way to obtain WKB actions.

The WKB actions we computed for both background fields yields compatible results with their physical properties. In the magnetic case, we obtained a real contribution indicating no real particle creation and it is related to the analogous of Landau levels of the uniform magnetic fields. For the electric case, on the other hand, we obtained both real and imaginary contributions which arise from a \textit{linear combination} of two WKB cycles above the barrier top. The imaginary part matches with WLI action for the time dependent periodic electric field; therefore, it is associated to the particle creation\footnote{ Note that a similar observation was made in the direct quantum mechanical setting in \cite{Basar_2015} where authors associated the WLI action with the tunneling action around the barrier top while in our construction, the electric case is precisely linked to the linear combination of WKB cycles above the barrier top and the WLI action is given by the imaginary part of this combination.}. In order to have a complete picture, we also discussed the periodic electric background potential starting from the corresponding Klein-Gordon equation and show how exactly the linear combination of the WKB cycles above the barrier top arises in EWKB approach. In this way, we verified the computations in time dependent formulation based on the analytical continuation and also provided a precise example on the connection between WLI and EWKB method.

\vspace{0.75cm}
\noindent \textbf{\underline{Outline of the paper}}: In Section \ref{Section: PP_propblem}, we first reformulated the real time approach by providing an unambiguous definition of the pair production probability which is compatible with the time dependent scattering amplitude description we discussed above. This reformulation is based on the well-known fact that $i\ve$ prescription which is associated to the forward/backward time flow directions in the scattering process and puts the ideas in \cite{Chadha:1977my}, where only one time direction for the electric case was discussed, on a more rigorous basis. After discussing this reformulation, we present its application to both uniform electric and magnetic backgrounds in Section \ref{Section: PP_Uniform} and show that how the contributions emerging from the proper time integrals are consistent with the unitarity condition and properties of their associated vacua.

In Section \ref{Section: PP_perturbative}, we extend our discussion to arbitrary background gauge potentials in arbitrary dimension.  We analyze general space dependent or time dependent electric fields under very general assumptions and compute the first two orders for the non-perturbative pair production probabilities in \textit{locally constant field approximation} (LCFA) limit. 

In the second part of the paper, we focus on the connection to the semi-classical methods. At the end of Section \ref{Section: PP_perturbative}, we verify our results obtained by the perturbative approach to WLI instanton actions of the periodic and single-pulse electric field configurations \cite{Dunne:2005sx, Dunne:2006st}. Then, in Section \ref{Section: SC_Connection}, considering both electric and magnetic periodic backgrounds, we focus on the connection to the WKB approach. In Section \ref{Section: Compare_WKB}, we first show how the WKB loop integrals emerge in the real time formulation via the decomposition of the effective action which we explain in Section \ref{Section: PP_perturbative} and compute the corresponding actions via WKB integrals. Then, we provide an alternative computations by direct integrations along Lefschetz thimbles in Section \ref{Section: Thimbles}. Finally,  we finish with a discussion on the connection to EWKB method in Section \ref{Section: ConnectionEWKB}, where we obtain precise WKB cycles associated to the real time electric background computations.

\section{Pair Production Problem and Resolution of Ambiguity}\label{Section: PP_propblem}
In this paper, we  only consider the pair production of bosonic matter as the fermionic matter production can be handled equivalently. For bosonic matter fields in the presence of an electromagnetic background, the effective action at 1 loop order is expressed as
\begin{equation}\label{EffectiveAction_1Loop}
	\G(m^2) =  i\ln \det (m^2 + \mcalH),
\end{equation}
where $m^2$ is the mass of bosonic matter particles and 
\begin{equation}\label{Hamiltoian}
	\mcalH = \frac{1}{2}(p_\m - e A_\m)^2
\end{equation}
is the Hamiltonian operator governing the motion of the bosonic particles. Note that in this paper, we use $g_{\m\n} = \diag(- + + +)$ metric convention. 

The effective action $\G(m^2)$ is related to the Green's operator $G^\pm(z)$ as
\begin{equation}
	\Tr G^\pm(m^2) = \frac{\mrmd \G^\pm(m^2)}{\mrmd m^2} ,
\end{equation}
which can also be written by an inverse derivative as
\begin{equation}\label{ActionFromResolvent}
	\G^\pm(m^2) = \int^{m^2}\mrmd z\,\Tr G^\pm(z),
\end{equation}
Integrating over singularities of $G^\pm(z)$ induces a discontinuity for the effective action \cite{Eden:1966dnq} and we define a function in $\ve \rightarrow 0$ limit to represent this discontinuity
\begin{equation}\label{PP_gap}
	\D \G = \G^+(m^2) - \G^-(m^2)
\end{equation}
whose imaginary part, i.e. $\Im \D\G(m^2)$, is the total pair production probability.

In the proper time dependent setting, the actions $\G^\pm$ are written by the Schwinger's integrals as
\begin{align}
	\G^\pm (m^2) & = \mp \int^{m^2} \mrmd z\, \int_0^\infty \mrmd t\, e^{\mp i  m^2 t}\,  \Tr e^{\mp i H t}, \label{SchwingerIntegral_Pre}\\
	&= -i \int_{J^\pm_0}  \frac{\mrmd t}{t} \, e^{\mp i  m^2 t}\,  \Tr e^{\mp i H t},\label{SchwingerIntegral}
\end{align}
where we analytically continued the integration contour rather than using $i\ve$ prescription and defined $J_0^\pm = \left[0,\infty e^{\mp i \t}\right]$. This is equivalent to $i\ve$ prescription and we use both analytical continuations in this paper when they are needed. Note that we omitted $\t\rightarrow 0^+$ limit in \eqref{SchwingerIntegral} for notational simplicity and we will continue using this notation throughout the paper.

In \eqref{SchwingerIntegral}, $t$ integral can be considered as a Borel-like integral. The integrand has possible singularities at $t\neq 0$ containing non-perturbative information which are the pair production probabilities in our context. Contrary to the conventional Borel integral, on the other hand, the analytical continuation directions are pre-determined. Thus, there is no Borel-like ambiguity takes place for \eqref{SchwingerIntegral}.

Note that there is also another pole at $t=0$, which corresponds to the UV divergence in quantum field theories. In 4 dimensional QED, $\Tr e^{\mp i H t}$ has another singularity at $t=0$ which is handled by the renormalization procedure but we don't discuss them in this paper.

\paragraph{\underline{Remark}:} $\Tr e^{\mp i H t}$ is the proper time evolution operator for an observer in particle's rest frame as $H$ generates translations in the proper time $t$. Then, $\G^+$ and $\G^-$ correspond to effective actions for forward and
backward proper-time evolutions respectively. Let us make some formal manipulations and rewrite the latter one by setting $t\rightarrow -t$:
\begin{equation}
	\G^-(m^2) = i \int_{-\infty}^0 \frac{\mrmd t}{t}\, e^{-i m^2 t} \, \Tr e^{-i H t}.
\end{equation}
where $\t\rightarrow 0^+$ limit is taken. Now, the action is described by forward time evolution between $t=-\infty$ and $t=0$. Moreover, in this form the gap equation in \eqref{PP_gap} becomes 
\begin{equation}\label{PP_gap_fullevolution}
	\D \G(m^2) = -i \int_{-\infty}^{\infty}\frac{\mrmd t}{t}\, e^{-im^2t}\, \Tr e^{-i H t}
\end{equation}
which defines the effective action of a propagation between infinite past and future of an observer in particles' rest frame. Then, from \eqref{PP_gap_fullevolution} we observe that the unambiguous definition in \eqref{PP_gap} assumes the proper time flow direction from past infinity to future infinity. In this sense, $\Im\D\G$ is the pair production probability with correct sign and the unitarity condition demands $\Im \D\G > 0$.

\subsection{Ambiguity and its Resolution in Uniform Electromagnetic Background}\label{Section: PP_Uniform}

Before considering general electromagnetic backgrounds, we are going to discuss how the unambigous pair production rate emerges from the exact effective action in 
both uniform electric and magnetic backgrounds and precisely show that the real proper time approach leads to unambiguous results for both cases, which are consistent with the unitarity condition and very well known properties of the vacua of these backgrounds.

\begin{figure}
	\centering
 	\includegraphics[width=0.75\textwidth]{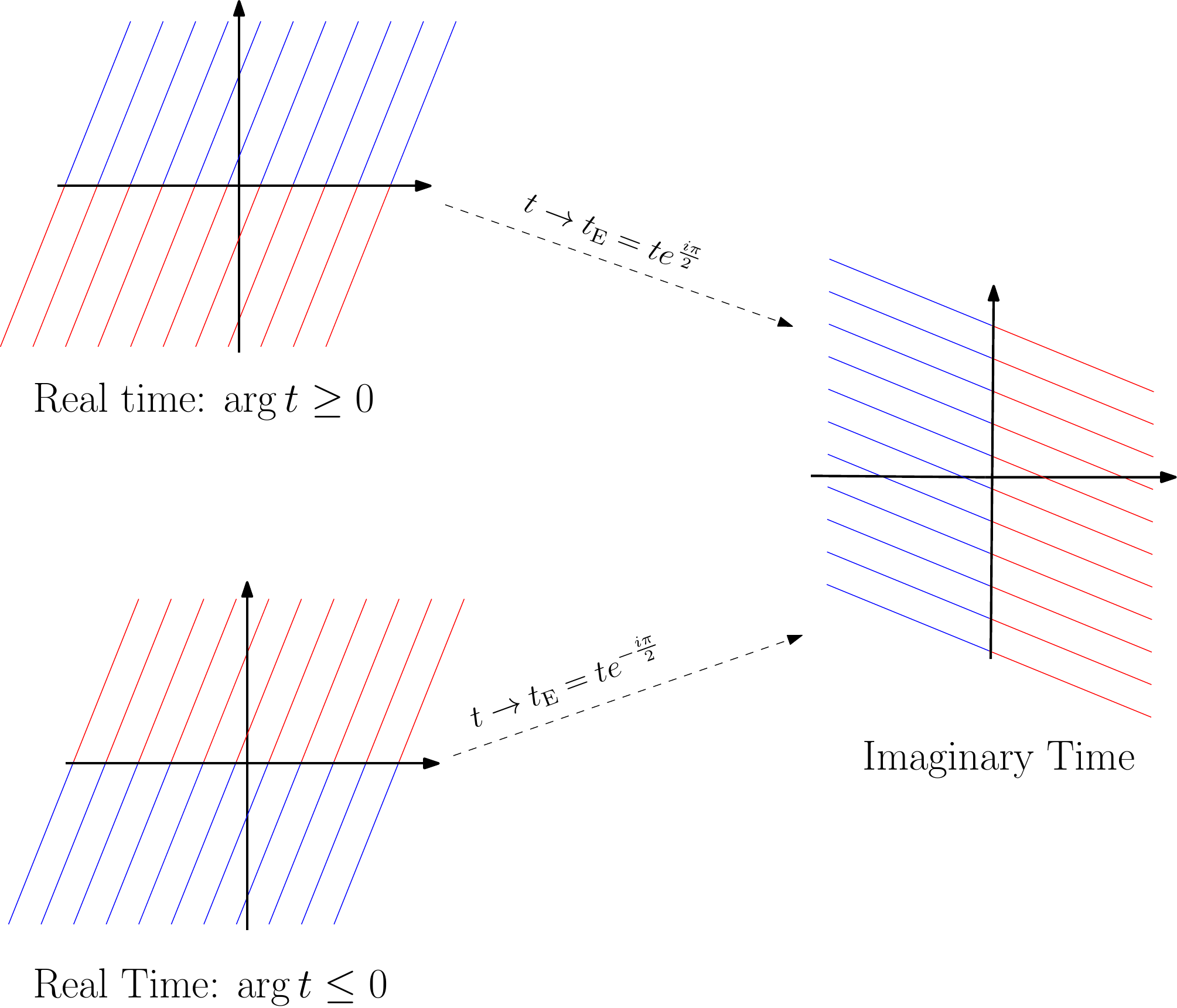}
	\caption{Complex $t$ planes. (Left) \underline{Real time cases}: Regions allowed for analytical continuations are represented by \textcolor{blue}{blue} and disallowed regions are represented by \textcolor{red}{red}. (Right) \underline{Imaginary time case}: This is obtained by rotating real time ones by $\frac{\pi}{2}$ in the respective allowed directions. Note that \textcolor{blue}{blue} and \textcolor{red}{red} regions on the right correspond to the same regions on the right.    }
	\label{Figure: WickRotations}
\end{figure}

For completeness, let us first look at the imaginary time case and review how the Borel ambiguity arises. The 1 loop effective action in Euclidean proper time for the bosonic fields in a uniform electromagnetic background is written as
\begin{equation}\label{SchwingerBosonic_Euclidean}
	\G_\mathrm{Eucl}(m) = \frac{\mcalV_4}{16\pi^2}\int_{0}^{\infty} \frac{\mrmd s}{s^3}\, \frac{e^{-  s m^2} (s e \E) (s e \B) }{\sin\left(\frac{s e \E}{2}\right)\sinh\left(\frac{s e \B}{2}\right)}.
\end{equation}
where $\mcalV_4 = L^3 T$ is the space-time volume factor. 

The main problem here is the loss of the information provided by $i\ve$ prescription. In fact, when we start with the Euclidean time, the exponential term makes the integral converging as the contour approaches to infinity. However, when there is a singularity on the contour, there is no guidance stating how to handle it. This is illustrated in Fig.~\ref{Figure: WickRotations} where the allowed/forbidden regions for forward and backward time evolutions are merged after proper Wick rotations. Therefore, the distinction between forward and backward time evolutions vanishes and the expression \eqref{SchwingerBosonic_Euclidean} represents the Wick rotated effective action for both cases. This also prevents from defining the gap $\D\G$ unambiguously as in \eqref{PP_gap}. This also shows the origin of the ambiguity in the Euclidean case and why its resolution needs a real time approach.

We first review the standard picture for the pair production to emphasize its problem. In \eqref{SchwingerBosonic_Euclidean}, the poles related to the background electric field lie on the real axis at $t=\pm \frac{n \pi}{e \E}$ while the poles corresponding to magnetic background are on the imaginary axis $t = \pm i \frac{n \pi}{e \B}$, where $n\in \mbbN^+$ for both cases. It is possible to compute the imaginary contributions using standard contour integration. Common lore is that the electric poles lead to emergence of $\Im\G$ while the magnetic ones are not relevant since they do not lie on the integration contour in \eqref{SchwingerBosonic_Euclidean}. 

\begin{figure}
	\centering
	\includegraphics[width=0.6\textwidth]{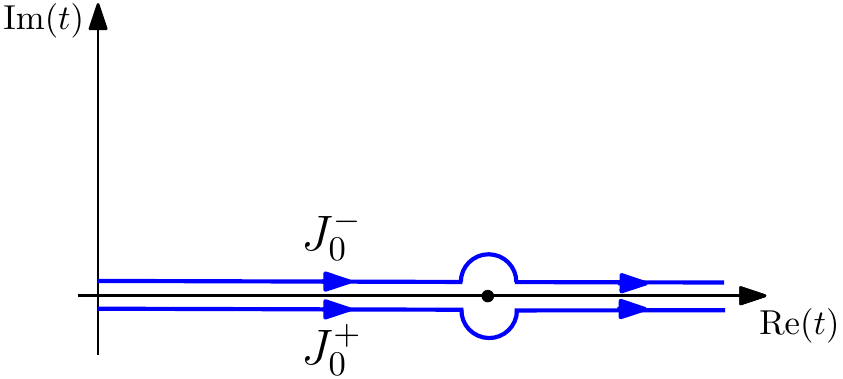}
	\caption{ Possible Borel integral contours on the real axis for uniform electric field background in the Euclidean proper time and uniform magnetic background in the real time case.}
	\label{Figure: EuclideanBorel}
\end{figure}

Consider the contribution of the first electric pole. There are two possible ways to analytically continue which leads to two distinct integration paths, i.e. $J_0^+$ and $J_0^-$. (See Fig \ref{Figure: EuclideanBorel}). There is no guideline for choosing any of these paths; therefore we have two complex conjugate results: 
\begin{equation}\label{PP_Euclidean}
	\Im \G \sim \mp  e^{-\frac{m^2 \pi}{e \E}} .
\end{equation}
We know that the pair production probability is defined by $\mcalP \sim 2\Im \G $ and can only be positive. Therefore, we can just choose the result with $+$ sign in \eqref{PP_Euclidean}. However, this is an ad-hoc approach and does not hide the fact that the result is ambiguous. This is equivalent to the Borel ambiguity and as we mentioned it can be overcome by real proper time approach which provides us an ambiguous definition of the pair production probability as we described above.

Let us now return to the real proper time case. The effective action in real time is 
\begin{equation}\label{SchwingerBosonic_Full}
	\G^\pm(m) = -\frac{\mcalV_4}{16\pi^2}\int_{J^\pm_0} \frac{\mrmd t}{t^3}\, \frac{e^{\mp i t m^2} (t e \E) (t e \B) }{\sinh\left(\frac{t e \E}{2}\right)\sin\left(\frac{t e \B}{2}\right)}.
\end{equation}
Note that the location of the poles related to electric and magnetic fields exchanged and for both $\G^\pm$ the possible analytical continuation directions are determined by definition. In the following, for simplicity of discussion, we continue with pure electric field and pure magnetic field backgrounds.

\paragraph{Uniform Electric Background:} In $\B\rightarrow 0$ limit, the effective action \eqref{SchwingerBosonic_Full} becomes
\begin{equation}\label{SchwingerBosonic_Electric}
	\G_\mrmE^{\pm}(m) = -\frac{\mcalV_4}{4\pi^3}\int_{J^\pm_0} \frac{\mrmd t}{t^3}\,  \frac{e^{\mp i t m^2} (t e \E)}{2\sinh(\frac{t e \E}{2})}
\end{equation}
and the integrand has poles at \[t_\mrmE = \pm \frac{2\pi i n_\mrmE }{e \E} \quad , \quad n_E \in \mbbN^+ . \] While all the poles can be treated collectively, the pair production rate of first particle and anti-particle pair is only linked to the first order non-perturbative term \cite{Cohen:2008wz}. Therefore, only this part needs to satisfy the unitarity condition without any ambiguity and we will consider the first poles at $t_\mrmE = \pm\frac{2\pi i}{e\E} $ in our analysis. 

Both $\G^+_\mrmE$ and $\G^-_\mrmE$ contain poles at $\pm\frac{2\pi i}{e\E} $ in the complex $t$ plane. However, since the analytical continuation of $t$ is allowed only in one direction by construction, $\G^+_\mrmE$ and $\G^-_\mrmE$ can only get contributions from one of the poles, i.e., $-\frac{2\pi i}{e\E}$ or $+\frac{2\pi i}{e\E}$ respectively.

\begin{figure}
	\centering
	\begin{subfigure}[b]{0.4\textwidth}
		\includegraphics[width=\textwidth]{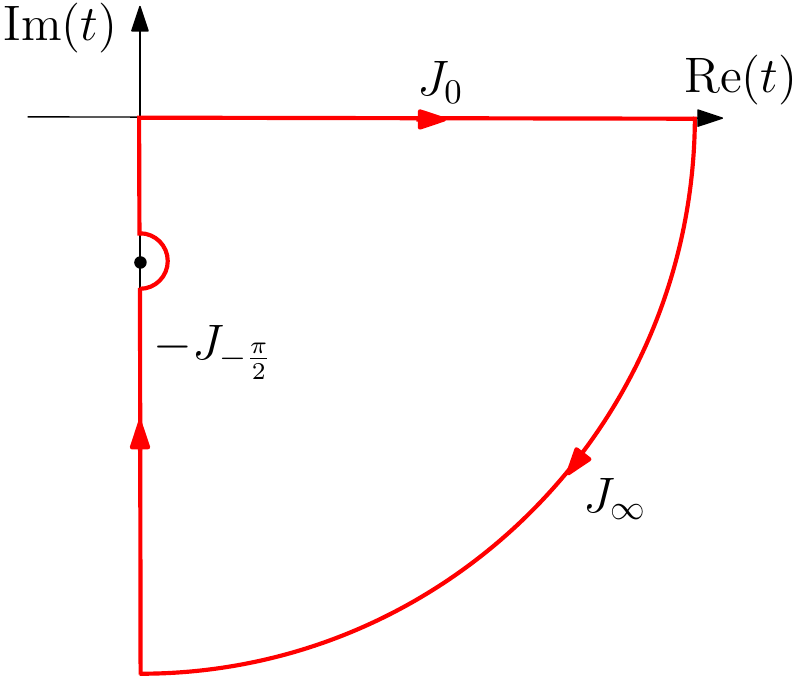}
		\caption{}
		\label{Figure: ConstantElectricContour2}
	\end{subfigure}
	~ 
	\begin{subfigure}[b]{0.4\textwidth}
		\includegraphics[width=\textwidth]{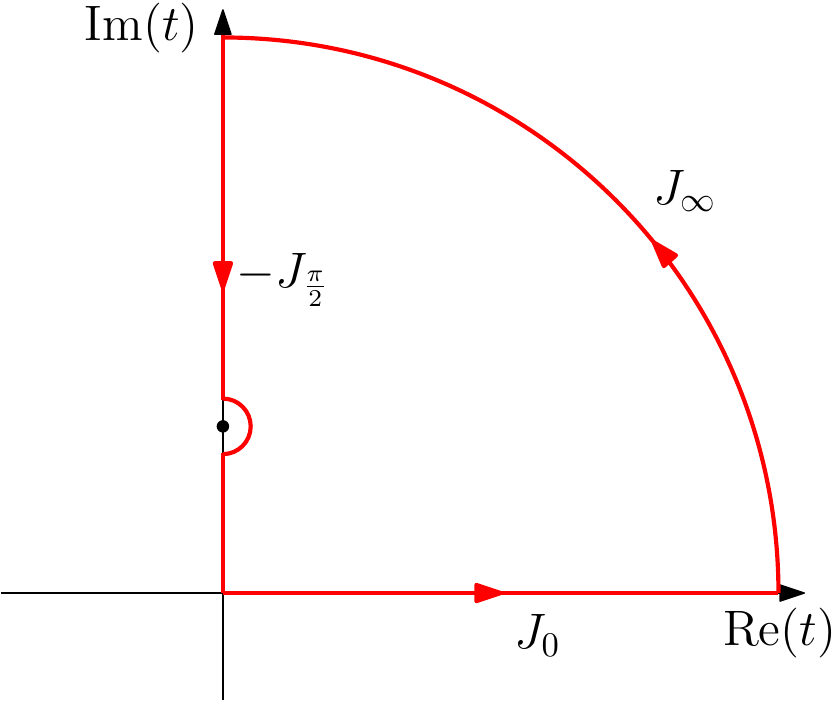}
		\caption{}
		\label{Figure: ConstantElectricContour}
	\end{subfigure}
	~ 
	\caption{Contours for the real time integrals in Electric case. \ref{Figure: ConstantElectricContour2}: Contours for $\G^+_\mrmE$. \ref{Figure: ConstantElectricContour}: Contours for $\G^-_\mrmE$. }\label{Figure: ElectricContours_constant}
\end{figure}

First, we consider $\G^+_\mrmE(m)$. To get the contribution from the pole at $-\frac{2\pi i}{e \E}$, we simply rotate in clockwise direction by $\frac{\pi}{2}$ degree as in Fig. \ref{Figure: ConstantElectricContour} and using $\int_{J_0} + \int_{-J_{-\pi/2}^+} = 0$, we re-write the effective action as 
\begin{align}
	\G^+_\mrmE(m^2) &= - \mcalV_4\frac{e\E}{8\pi^2} \int_{J^+_{-\frac{\pi}{2}}} \frac{\mrmd t}{t^2} \frac{e^{-im^2 t}}{\sinh\left(\frac{t e\E}{2}\right)} \label{RotatedElectric_+} 
\end{align}
Note that the analytical continuation of the contour $J_{\pi/2}^-$ around the singularity at $t= -\frac{2\pi i}{e\E}$ is fixed by the rotation direction of the initial contour on the real axis. Then, we obtain the imaginary part of $\G^+_\mrmE$ with a fixed sign as
\begin{equation}
	\Im \G^+_\mrmE(m^2) = + \mcalV_4 \frac{(e\E)^2}{16\pi^2}e^{-\frac{2\pi m^2}{e\E}}.
\end{equation}
In the same way, we can extract the imaginary part of $\G^-_\mrmE(m^2)$ from the pole at $t= \frac{2\pi i}{e\E}$. This time, we rotate the contour in counter-clockwise direction as in Fig. \ref{Figure: ConstantElectricContour2} and re-write the effective action as
\begin{align}
	\G^-_\mrmE(m^2) & =  - \mcalV_4\frac{e\E}{8\pi^2} \int_{J^-_{\frac{\pi}{2}}}  \frac{\mrmd t}{t^2}\frac{e^{i t m^2}}{\sinh( \frac{t e \E}{2})} \label{RotatedElectric_-}
\end{align}
and get the imaginary part of $\G^-_\mrmE$ as
\begin{equation}
	\Im \G^-_\mrmE(m^2) = - \mcalV_4 \frac{(e\E)^2}{16\pi^2}e^{-\frac{2\pi m^2}{e\E}}.
\end{equation}
Combining these two results, we get the imaginary part of the gap $\D \G = \G^+ - \G^-$, so that the pair production rate as
\begin{equation}\label{PP_ConstantElectric}
	\mcalP = \Im \D \G_\mrmE(m^2) = \mcalV_4 \frac{(e\E)^2}{8\pi^2}e^{-\frac{2\pi m^2}{eE}},
\end{equation}
which recovers the standard pair production probability with an unambigous sign. Note that along with the individual results for $\Im \G^\pm$, the unambigous definition of $\D\G$ is also a primary factor for the unambiguous end result. Without this definition, the ambiguity would be persistent in the real time case in the same way with the Euclidean case.



\paragraph{Uniform Magnetic Background:}
Although the uniform magnetic case does not yield pair production, it is important to see how the unambiguous definition of $\D \G$ fits in this case. In $\E\rightarrow 0$ limit, \eqref{SchwingerBosonic_Full} reduces to
\begin{equation}
	\G^\pm_\mrmB(m^2) = -\frac{\mcalV_4}{4\pi^2}\int_{J^\pm_0} \frac{\mrmd t}{t^3}\frac{e^{\mp i t m^2} t e B}{2\sin\left(\frac{t eB}{2}\right)}.
\end{equation}
Now the poles are on the real axis and the treatment is similar to the standard Borel integral. However, again as in the electric background case, the restriction on the analytical continuation directions for both $\G^+_\mrmB$ and $\G^-_\mrmB$ prevents ambiguous results. Taking the integrals using the appropriate contours as pictured in Fig. \ref{Figure: EuclideanBorel}, we find the gap $\D \G_\mrmB = \G^+_\mrmB- \G^-_\mrmB$ due to the  pole at $t_\mrmB = \frac{2\pi}{e \B}$ as 
\begin{equation}\label{PP_ConstantMagnetic}
	\D \G_\mrmB = -\frac{eB}{8\pi^2} \left[ \left(-i \pi\frac{eB}{2\pi}\right)e^{-\frac{2 \pi i m^2}{eB}} - \left(i \pi\frac{eB}{2\pi}\right)e^{\frac{2\pi i m^2}{eB}}   \right] =  i \frac{(eB)^2}{8\pi^3}\cos\left(\frac{2\pi m^2 }{eB}\right)
\end{equation}
This is again imaginary contribution with a sign that keeps the theory unitary but now it has an oscillatory behaviour. This is fine as the it indicates that the stable quantum vacuum is not completely empty and there are successive particle creations and annihilations as expected from any relativistic quantum theory but effectively this process leads to no effective pair creation. Therefore, there is no need for any cancellation of this contribution by Bogomolny Zinn-Justin mechanism.

\section{Unambiguous Pair Production from Perturbative Expansions}\label{Section: PP_perturbative}

In physics, exact solutions are very rare. Therefore, while the discussion in Section \ref{Section: PP_Uniform} explicitly show  how the pair production rate in the presence of a uniform electromagnetic field emerges in a way that the unitarity is preserved, it is also important to show how the same information can be achieved when an exact solution is not accessible. 

In absence of an exact solution, perturbative techniques allow us to compute the coefficients in a perturbation series, which form generically divergent series. The non-perturbative information is encoded in this divergent series and can be extracted using Borel-Pade techniques \cite{Costin:2019xql,Costin:2020hwg,Costin:2020pcj,Florio:2019hzn}. However, the Borel-Pade summation leads to the same ambiguity that we discussed in the Euclidean case in the previous section and whenever there is a persistent instability of the vacuum. 
As in the uniform case, to get an unambiguous result, we use the real time approach and probe the poles of the propagator directly, i.e. we need to sum the perturbative expansion of the propagator $\Tr U^\pm(t) = \Tr e^{\mp i t H}$ in \eqref{SchwingerIntegral} before taking the proper time integral. 

For this reason, we adapt the recursive perturbative scheme introduced in \cite{Pazarbasi:2020god} which is based on the construction in \cite{Moss:1993rc,Moss_1999} to problems with background gauge potentials.  In \cite{Pazarbasi:2020god}, it was shown that the non-commutativity of phase-space variables acts as a source for a derivative expansion which is identified as the semi-classical expansion. In the present context, the non-commutativity again produces a derivative expansion which corresponds to the deviation from the uniform field. At each order of the derivative expansion, there is also another expansion which depends on the electromagnetic field strength and using this one the non-perturbative pair production probability at each order in the derivative expansion can be obtained. 

With the perturbative scheme that we will explain shortly, we compute the pair production probability in arbitrary temporal or spatial electric field backgrounds up to the $2^\nd$ order in derivative expansion. Throughout this section we rely on the perturbative expansions and compute the non-perturbative results via the summation of the series expansions. Our series expansions are compatible with the exact results found in \cite{Gusynin:1995bc,Gusynin:1998bt} but since our approach relies on purely perturbative calculations and free from the complications of the exact computations. In addition to that the pair production probability we compute at the $2^\nd$ order slightly differs from \cite{Gusynin:1998bt} in presence of spatial inhomogeneities.


Note that the non-perturbative pair production problem based on the summation of the perturbative expansions has been discussed for an exactly solvable one dimensional electric background in \cite{Dunne:1999uy,Dunne:2022esi}. The content of this section can be seen as a first step of the generalization of these results to general background potentials in arbitrary dimensions while keeping the unitarity of the physical quantities without any ad-hoc choices. 

\vspace{0.5cm}
\textit{\underline{Recursive Expansion}:}
In this section, we will investigate pure electric fields. The Hamiltonian is written in general as
\begin{equation}\label{Hamiltonian_BackgroudEM}
	H = -\frac{1}{2}(p_0 - e A_0(\mbfx))^2 + \frac{1}{2}( \mbfp- e \mbfA(x_0))^2 \equiv -\frac{1}{2}\pi^2_0(p_0,\mbfx) +\frac{1}{2}\mbfpi^2(\mbfp ,x_0).
\end{equation}
We assume that the time like and space like parts of the gauge potential are in the following forms respectively:
\begin{equation}\label{GaugePotential}
	A_0(x_0,\mbfx) = -\frac{\E}{\o}H_0(\o\, \mbfx) \quad , \quad \mbfA = -\frac{\E}{\o}\mbfH(\o x_0).
\end{equation} 
Then, the electric field is 
\[\mbfE = \frac{\E}{\o} \left( \frac{\mrmd \mbfH(\o x_0)}{\mrmd x_0} +  \nabla H_0(\o \, \mbfx)\right).\] Since we are interested in the locally uniform fields, we assume that both $H_0(\o \mbfx)$ and $\mbfH(\o x_0)$ are slowly varying functions of $\mbfx$ and $x_0$ respectively. For the same reason, we also assume that $m \gg e \E \gg \o $ throughout this paper. This will guide us in our calculations.

The effective action for the general Hamiltonian in \eqref{Hamiltonian_BackgroudEM} is written as
\begin{equation}\label{effectiveAction_pureElectric_Reduced}
	\G^\pm(m^2)  = - i \int_{J^\pm_0} \frac{\mrmd t}{t}e^{\mp i m^2 t} \int\frac{\mrmd^4 p}{(2\pi)^4}\, \<p_0,\mbfp |e^{\mp i t \left(-\frac{1}{2}\pi_0^2(p_0,\mbfx) + \frac{1}{2}\mbfpi^2(\mbfp,x_0)  \right) } |\mbfp, p_0 \>.
\end{equation}
Note that since $\pi_0$ and $\mbfpi$ do not commute with each other, the propagator does not factorize trivially. To get the perturbative expansion, we can either redefine the propagator $	U(t) = \Tr e^{\mp i t H} $ as
\begin{equation}\label{choice1}
	U(t) = e^{\pm \frac{i t \pi^2_0}{2}} \tU(t)
\end{equation}
or 
\begin{equation}\label{choice2}
	U(t) = e^{\mp \frac{i t \mbfpi^2}{2}} \tU(t).
\end{equation}
As shown in \cite{Pazarbasi:2020god}, both choices are equivalent. Solving the time dependent Schrodinger equation for redefined propagators and following the steps in \cite{Pazarbasi:2020god}, we reach a recursive expansion for $\tU(t)$ for each case and express the effective action in terms of this recursive expansion. 

Let us consider pure time dependent and pure space dependent electric fields. For the time dependent case, where  $A_0(\mbfx) = 0$ and $\pi_0 = p_0$, the perturbative expansion is written as 
\begin{align}\label{QuantumAction_timeDependent}
	\G^\pm_{\mathrm{time}} & = -i\sum_{m=0}  \int_{J^\pm_0} \frac{\mrmd t}{t} e^{\mp im^2t} \int  \frac{\mrmd^4 p\, \mrmd^4 x }{(2\pi)^4}\,e^{\mp  \frac{i\mbfpi^2 t}{2}}  \,  \sum_{k=1}^{m} \tU^\pm_{m,k}(t)  \, e^{\pm \frac{i p_0^2 t}{2}} ,
\end{align}
where $\tU^\pm$ is given by the recursion relation
\begin{align}
	\tU_{m,k}^\pm(t) & =\frac{1}{2}\sum_{l=1}^{m-k+1}\frac{1}{l!}\frac{\dee^l \mbfpi^2} {\dee x_0^l} \int_{0}^t \mrmd t_1\, \,  \msfb_\mp^l(p_0,\dee_{p_0},t_{1})\tU_{m-l,k-1}^\pm(t_1) , \label{RecursionRelation_1}
\end{align} 
with the initial value $\tU^\pm_{0,0} = e^{\pm \frac{i p_0^2 t}{2}}$ and the operator valued function
\begin{equation}\label{operator1}
	\msfb_\mp(p_0,\dee_{p_0}, t_i)=i \dee_{p_0}  \mp p_0\, t_i \pm p_0\, t .
\end{equation}
Note that equations \eqref{QuantumAction_timeDependent} - \eqref{operator1} indicate that the problem is effectively one dimensional regardless the details of the background gauge field. 

If the background field, on the other hand, is space dependent, then the problem might be a multi-dimensional one. In this case, $\mbfA(x_0)=0$ , $\mbfpi = \mbfp$ and the perturbative expansion becomes
\begin{equation}\label{QuantumAction_spaceDependent}
	\G^\pm_{\mathrm{space}}  = -i\sum_{m=0}  \int_{J^\pm_0} \frac{\mrmd t}{t} e^{\mp im^2t} \int  \frac{\mrmd^4 p\, \mrmd^4 x }{(2\pi)^4}\,e^{\pm \frac{i \pi_0 t}{2}}  \,  \sum_{k=1}^{m} \tU^\pm_{m,k}(t)  \, e^{\mp \frac{i \mbfp^2 t}{2}} ,
\end{equation}
where $\tU^\pm$ is 
\begin{align}
	\tU_{m,k}^\pm(t) & =\frac{1}{2}\sum_{l=1}^{m-k+1}\frac{1}{l!} \nabla^l \pi_0^2 \int_{0}^t \mrmd t_1\, \,  \msfb_\pm^l(\mbfp,\nabla_{\mbfp},t_{1})\, \tU_{m-l,k-1}^\pm(t_1)  , \label{RecursionRelation_2}
\end{align} 
with $\tU_{0,0}^\pm = e^{\mp \frac{i \mbfp^2 t}{2}}$ and
\begin{equation}\label{operator2}
	\msfb_\pm(\mbfp,\nabla_{\mbfp}, t_i)=i \nabla_{\mbfp}  \pm \mbfp\, t_i \mp \mbfp\, t .
\end{equation}

\vspace{0.5cm}
\textit{\underline{Structure of the Perturbative Expansions}:} The expressions \eqref{RecursionRelation_1} and \eqref{RecursionRelation_2} shows that each expansion is organized as a derivative expansion on $\mbfpi^2$ and $\pi^2_0$. This suggests that the perturbative expansions of the proper-time propagator in a double expansion in $\o$ and $e \mcalE$ as
\begin{equation}\label{generalDoubleExpansion}
	\frac{\Tr e^{\mp i t H}}{t} = \sum_{m,n} \a_{m,n}(t) \,  \o^{2m} (e\mcalE)^{2n}.
\end{equation}
Note that in \eqref{generalDoubleExpansion}, the order of $\o$ expansion reflects the number of derivatives applied to the gauge potentials $A_0(\mbfx)$ or $\mbfA(x_0)$ at a given order in $e\E$ expansion, while the order of $e\E$ expansion is equal to the number of gauge potentials at that order. Then, for both time dependent and space dependent backgrounds, at a fixed order in $\o$, the expansion in $e\E$ can be organized such that the series consists of $\left(e\E\dee_0 \mbfA(x_0)\right)^2$ and $\left(e\E\nabla A(\mbfx)\right)^2$ terms respectively, while higher order derivatives on the background field appear as pre-factors to these series.

We determine the per-factors for the first two order in $\o$ expansion as in the following. First observe that since the coefficients $\a_{m,n}$ are computed upon integrating the series, which obtained recursively via \eqref{RecursionRelation_1} or \eqref{RecursionRelation_2}, over 4-momenta $(p_0,\mbfp)$, all the terms with odd powers $p_0$ or $\mbfp$ will vanish. For example, $\tU_{1,1}^\pm$ term for the time dependent case reads from \eqref{RecursionRelation_1} as 
\begin{align}
	\tU_{1,1}^\pm(t) &=  \frac{\dee_0 \mbfpi^2}{2} \int_{0}^t \mrmd t_1\, \msfb_\mp(p_0,\dee_{p_0},t_1)\, e^{\pm \frac{i p_0^2 t}{2} } \cr
	& =  \mp \frac{p_0 t^2}{2}\left(\mbfp - \mbfA(x_0)\right) \dee_0\mbfA(x_0) \, e^{\pm \frac{i p_0^2 t}{2}} \label{U11_time}
\end{align}
which is linear in both $p_0$ and $\mbfp$ and through the 4-momentum integration its contribution to \eqref{QuantumAction_timeDependent} would vanish. Similarly for space dependent case, we get
\begin{align}
	\tU_{1,1}^\pm(t) &=  \frac{\nabla \pi_0^2}{2} \int_{0}^t \mrmd t_1\, \msfb_\pm(\mbfp,\nabla_\mbfp,t_1)\, e^{\mp \frac{i \mbfp^2 t}{2} } \cr
	& = \pm \frac{\mbfp \,t^2}{2} \left(p_0 - A_0(\mbfx)\right)\nabla A_0(\mbfx)\, e^{\mp \frac{i \mbfp^2 t}{2}} \label{U11_space}
\end{align}
The first non-trivial contribution arises from $\tU_{2,1}^\pm(t)$ term. It is written as 
\begin{align}
	\tU_{2,1}^\pm(t)  &= \frac{\dee_0^2 \mbfpi^2}{2} \int_0^t \mrmd t_1\, \msfb_\mp^2(p_0,\dee_{p_0},t_1) e^{\pm\frac{ip_0^2 t}{2}} \cr
	& = \left(\dee_0\mbfA \cdot \dee_0\mbfA - (\mbfp - \mbfA)\dee^2\mbfA\right) Q^\pm_{\mathrm{time}}(p_0,t)\label{U21_time}
\end{align}
for the time dependent case and 
\begin{align}
	\tU_{2,1}^{\pm}(t) & = \frac{\nabla^2 \pi_0^2}{2} \int_0^t \mrmd t_1\, \msfb_\pm^2(\mbfp,\nabla_\mbfp,t_1) e^{\mp\frac{i\mbfp^2 t}{2}} \cr
	& = \left(\nabla A_0 \cdot  \nabla A_0 - (p_0 - A_0)\nabla^2 A_0\right) Q^\pm_{\mathrm{space}}(\mbfp,t)\label{U21_space}
\end{align}
for the space dependent case. For both cases, only the first terms contribute to the effective action \eqref{QuantumAction_timeDependent} and \eqref{QuantumAction_spaceDependent}, while the second terms vanish upon momentum integrals and $\tU_{2,1}^\pm$ contributes only to the leading order in LCFA, i.e. $\o^0$. Higher order terms in $e\E$ expansion can be obtained in 	 a similar manner and throughout the series all the space dependence manifests itself as $|\dee_0\mbfA|^{2n}$ and $|\nabla A_0|^{2n}$ while odd terms vanishes through momentum integrals as in \eqref{U11_time} and \eqref{U11_space}. 

First non-vanishing corrections to the leading order expansion arises from $\tU_{4,2}^\pm$ term as
\begin{align}
	\tU_{4,2}^\pm(t) &= \sum_{l=1}^{3}\dee^l_0 \mbfpi^2 \int_0^t \mrmd t_1\, \msfb^l_\mp(p_0,\dee_{p_0},t_1) \,U_{3,l}\cr
	& = \sum_{l=0}^2 \left(\dee_0^{l+1}\mbfpi^2 \cdot \dee_0^{3-l}\mbfpi^2 \right) Q^{\pm,(l)}_{\mathrm{time}}(p_0,t)\label{U42_time}
\end{align}
for time dependent fields and
\begin{align}
	\tU_{4,2}^\pm(t) &= \sum_{l=1}^{3}\nabla^l \pi_0^2 \int_0^t \mrmd t_1\, \msfb^l_\pm(\mbfp,\nabla_{\mbfp},t_1) \,U_{3,l}\cr
	& = \sum_{l=0}^2 \left(\nabla^{l+1}\pi_0^2 \cdot \nabla^{3-l}\pi_0^2 \right) Q^{\pm,(l)}_{\mathrm{space}}(\mbfp,t)\label{U42_space}
\end{align}
for space dependent fields.

The spatial dependence in \eqref{U42_time} and \eqref{U42_space} appears as $(\dee_0\mbfA \cdot \dee_0^3\mbfA) $, $(\dee_0^2\mbfA\cdot\dee_0^2\mbfA)$  and  $(\nabla A_0\cdot \nabla^3 A_0)$ and $(\nabla^2 A_0 \cdot \nabla^2 A_0)$ respectively. At order $\o^2$, all the higher order terms in $e\E$ expansion contain these terms along with the terms appear at $\o^0$ order, i.e. $|\dee_0 \mbfA|^{2n}$ and $|\nabla A_0|^{2n}$. This will help us to organized the series expansions before performing the Pade summation.


\vspace{0.4cm}
\textit{\underline{Phase Space Integrals}:}
Before summing the series at each $\o$ order, we first take the momentum integrals. At this point, it is appropriate to shift the relevant parts of the canonical momentum as $\mbfp \rightarrow \mbfp + \mbfA(x_0)$ in the temporal case and $p_0 \rightarrow p_0 + A_0(\mbfx)$ in the spatial case. In this way, we eliminate all the terms with momentum and spatial variables which are coupled to each other after the initial expansion. Then, all momentum integrals become Gaussian and it is straightforward to compute them order by order in \eqref{QuantumAction_timeDependent} and \eqref{QuantumAction_spaceDependent}. Note that each Gaussian integral will have the same pre-factor: 
\begin{equation}\label{PreFactors}
	\frac{1}{\left(\sqrt{\pm 2\pi i t}\right)^3\sqrt{\mp 2\pi i t}} = \pm \frac{1}{4\pi^2 i t^2},
\end{equation} where $(+)$ sign is for the forward proper-time evolution, while $(-)$ sign is for the backward evolution.  This is different than the standard prefactor of $4$ dimensional Gaussian integrals, i.e. $\left(\frac{1}{\sqrt{\pm 2\pi i t^2}}\right)^4 = -\frac{1}{4\pi^2 t^2}$. The difference is due to the sign change in the Lorentzian metric and overall imaginary factor is important as it contributes to the effective action directly.

\vspace{0.4cm}
\textit{\underline{Summation}:}
The poles at $t\neq 0$, which are the sources of the pair production probability as we discussed in the constant field case, are related to the summation of \eqref{generalDoubleExpansion}. In the limit $e \mcalE \gg \o$, which we are interested in this paper, it is convenient to sum $e\mcalE$ series at each order of the $\o$ expansion. Therefore, after the momentum integrals, we will sum $e\E$ expansion via Pade method for orders $O(\o^0)$ and $O(\o^2)$ separately to compute the imaginary contributions to the effective action which are the non-perturbative pair production probabilities at the corresponding orders up to the spatial integral which we will deal last.

\vspace{0.4cm}
\textit{\underline{Space-time Integrals}:} After the summation, the spatial integrals can be also computed perturbatively or, in some special cases, exactly when the gauge potentials $A_0(\mbfx)$ or $\mbfA(x_0)$ are known. In the remaining of this section, we will consider general gauge potentials for both temporal and spatial cases under some general assumptions which we will state in each case separately. Then, we will handle the spatial integrals via the saddle point approximation which will be possible due to the relation $m\gg e\mcalE \gg \o$. Note that in the temporal case, the spatial integral is one dimensional as $\mbfA(x_0)$ has one dimensional space-time dependence. On the other hand, the dimension $D$ of the space integral depends on the dimensionality of $A_0(\mbfx)$. Even though it does not change the exponential part of the pair production probability, it plays a role in prefactors. We will elaborate this, when we discuss the space dependent fields.

\subsection{Time Dependent Electric Fields:}

We first consider time dependent electric fields for 
\begin{equation}\label{GaugePotential_TimeDependent}
	\mbfA(x_0) = \frac{\E}{\o}\mbfH( \o x_0).
\end{equation}
As we stated above, this is effectively a one-dimensional problem and in principle it can be handled by exact WKB methods as in \cite{Taya:2020dco}. Here using the recursion relation and the Pade summation of perturbative expansion of the propagator, we will obtain \textit{unambiguous} version of the pair production probability. Our formulation also indicates a possible relationship between Schwinger's formalism and the WKB formulation. We will return this relation shortly at the end of this paper.

\vspace{0.4cm}
\textit{Notation:} As we mentioned before and it is apparent from \eqref{GaugePotential_TimeDependent}, the order in $\o$ expansion is equal to the number of derivative acting on $\mbfA(x_0)$. To follow the order of $\o$ properly, we introduce the following notation for the derivative of the gauge potential in \eqref{GaugePotential}:
\begin{equation}
	\frac{\dee^n \mbfA(x_0)}{\dee x_0^n} = \mcalE \o^{n-1} \mbfH^{(n)}(\o x_0)
\end{equation}
where with $\mbfH^{(n)}(\o x_0)$ we mean the vector corresponding to the $n^\th$ derivative evaluated at $x_0$.  Along with the assumption $m\gg e\E\gg\o$ , throughout this subsection, we will assume that $\mbfH^{(1)}(\o x_0)$ is slowly varying and it takes the following values at any of its saddle point which we denote as $x_0^{\{i\}}$:
\begin{equation}\label{Assumptions_time}
	i) \, \mbfH^{(1)}(\o x_0^{\{i\}}) = \mbfh^{\{i\}}_1 \quad , \quad ii) \, \mbfH^{(2)}(\o x_0^{\{i\}}) = 0 \quad, \quad iii) \, \mbfH^{(3)}(\o x_0^{\{i\}}) = -\o^2 \mbfh_3^{\{i\}} ,
\end{equation}
where both $\mbfh_1^{\{i\}}$ and $\mbfh_3^{\{i\}}$ are vectors evaluated at $x_0^{\{i\}}$
. Note that first two assumptions indicates the locally constant behaviour of $\mbfH(\o x_0)$ and states that the electric field is $\mbfE = \mcalE=\mathrm{constant}$  at the saddle points $x_0^{\{i\}}$, while third assumption implies the saddle points $x_0^{\{i\}}$ are non-degenerate.

\paragraph{Leading Order:}
Using the recursion formula \eqref{RecursionRelation_1}, we obtained high order expansion at order $O(\o^0)$ as
\begin{equation}\label{Expansion_TimeDependent_LO}
	\sum_{n=0}^\infty \a^\pm_{0,n} (t e\E |\mbfH^{(1)}|)^{2n}
\end{equation} 
where $|\mbfH^{(1)}| = \sqrt{\frac{\dee \mbfH}{\dee x_0} \cdot \frac{\dee \mbfH}{\dee x_0}}$. Before summing the series, it is convenient to  re-scale $t \rightarrow \frac{t}{e \E |\mbfH^{(1)}|}$ and express the effective action at order $O(\o^0)$, as
\begin{equation}\label{EffectiveAction_Expansion_ElectricFake}
	\G^\pm_{\mrmE,\o ^0} = - \mcalV_3 \frac{(e \E |\mbfH^{(1)}|)^2}{4\pi^2} \int_\mbbR \mrmd x_0\, \int_{J^\pm_0} \frac{\mrmd t}{t^3} \, e^{\mp i \frac{m^2 t}{e \E |\mbfH^{(1)}|} } \, \mcalS_{\o^0}^\pm(t)
\end{equation} 
where $\mcalV_3$ is the 3-dimensional volume associated to the integration over $\mbfx$ and 
\begin{equation}
	\mcalS_{\o^0}^\pm	= \sum_{n=0}^\infty \a^\pm_{0,n} t^{2n}
\end{equation}
is the series with the coefficients $\a_{0,n}^\pm$ being rational numbers. First 7 terms of the series are \begin{equation}\label{ExpansionCoefficients_ElectricFake_1D} 
	\left\{ 1, - \frac{ 1}{24 },\frac{7}{5760},-\frac{31}{967680}, \frac{127}{154828800}, -\frac{73}{3503554560}, \frac{1414477}{2678117105664000} \right\} 
\end{equation}
This expansion is equal to the expansion of the denominator of the integrand in \eqref{SchwingerBosonic_Electric} as expected. However, since the gauge potential is a space dependent function, the final expression for the pair production probability will have a small but significant difference with the uniform case probability in \eqref{PP_ConstantElectric}.


First two terms in the expansion in \eqref{EffectiveAction_Expansion_ElectricFake} are singular at $t=0$. They are the UV divergent terms and can be handled by renormalization methods. On the other hand, the series also contains  non-singular term  which are associated to singularities at $t\neq0$ upon their summation. Note that this series is a convergent one. It becomes divergent if we take the proper time integral directly. This is in fact the divergence predicted by Dyson in \cite{PhysRev.85.631} and it would need to be treated by the Borel procedure. Instead, similar to the uniform case in Section \ref{Section: PP_Uniform}, it is possible to probe the singularity structure represented by this series directly by using the Pade summation before taking the proper time integral. 

In computations of the pair-production probability, it is possible to ignore the UV divergent terms and perform Pade approximation when probing the poles at finite $t$. However, we observe that they affect the pre-factor as the $\frac{1}{t^3}$ term effects the residue around the singularities at $t\neq 0$. For this we keep them in the following calculations. As in the constant field case, a proper way to consider this residue problem would be regularizing the proper-time integral in \eqref{EffectiveAction_Expansion_ElectricFake} by a cut-off in its lower limit. Then, residues so that the pair-production probabilities can be computed without dealing with the UV divergence.

Computing the series $S_{\o^0}^\pm $  up to $O\left(t^{50}\right)$ and summing it via Pade method with precision $O(10^{-8})$ leads to 11 singularities for \textit{both} cases at
\begin{equation}
	t^\pm_\star = \pm 2 \pi i n \quad , \quad n ={1,\dots,11} .
\end{equation} 
which matches with the exact result. Note that for $n>11$ the locations of the singularities start to deviate from the exact result and needs high order term for better approximations.

As we indicated in the previous section, only the first pole is linked to the pair production probability. Therefore, we are only interested in the residue of the leading poles at $t^\pm_\star = \pm 2\pi i$. In addition to this, as in the uniform case, the sign of the imaginary parts are fixed by the analytical continuation directions of the proper-time contour and $\Im \G^+$ gets contribution only from $t_\star^-$ while $\Im \G^-$ arises from $t_\star^+$. The residues for both cases are the same:
\begin{equation}
	R^\pm_{\o^0} = 0.02533029 \simeq \frac{1}{4\pi^2}.
\end{equation}
and taking the analytical continuation directions into account, we find the imaginary part as
\begin{equation}\label{LC_integral_time}
	\Im \G^\pm_{\mrmE,\o^0} = \pm \mcalV_3\int_\mbbR \mrmd x_0 \, \frac{\left(e\E|\mbfH^{(1)}|\right)^2}{16\pi^3} \, e^{- \frac{2\pi m^2} {e \E |\mbfH^{(1)}|}} .
\end{equation}
This is the general expression of the leading order derivative expansion for a time dependent electric field and compatible with the general result in \cite{Gusynin:1995bc,Gusynin:1998bt}. The uniform case in Section \ref{Section: PP_Uniform}  is just the special case where $|\mbfH^{(1)}| =1$. 

The remaining integral in \eqref{LC_integral_time} can be evaluated exactly for specific cases. (See e.g. \cite{Dunne:1999uy}.) For general background fields, on the other hand, it is possible to handle the integral by saddle point approximation, which is appropriate since $m\gg e\E$ and $\mbfH(\o x_0)$ is chosen to be slowly varying.
Then, using the assumptions given in \eqref{Assumptions_time}, the leading order saddle point approximation leads to the leading order pair production probability as 
\begin{equation}\label{PP_time}
	\Im \D \G_{\mrmE,\o^0}(m^2) \simeq \mcalV_3 \sum_i \frac{(  e\E)^{5/2}\, |\mbfh^i_1|^3}{8 \pi^3 m\, \o}\sqrt{\frac{|\mbfh_1^{\{i\}}|}{\mbfh_1^{\{i\}}\cdot \mbfh_3^{\{i\}}} } \, e^{-\frac{2 \pi m^2}{e\E|\mbfh^{(i)}_1|}} ,
\end{equation}
where  $\mbfh^i_1$ and $\mbfh^i_3$ can be determined for specific background fields. Note that as we stated before, the prefactor in \eqref{PP_time} differs from the expression in uniform electric field case in \eqref{PP_ConstantElectric} and these two results are not equal to each other in $\o\rightarrow 0$, $\mbfh_1^{\{i\}}\rightarrow 0$ and $\mbfh_3^{\{i\}}\rightarrow 0$ limits as \eqref{PP_time} diverges in these limits. Instead, one should take the limit before handling the space integral in \eqref{LC_integral_time}.

\paragraph{Next to Leading Order:}
At the order $O(\o^2)$, the expansion \eqref{generalDoubleExpansion} can be organized in two parts: 
\begin{equation}\label{Expansion_TimeDependent_NLO1}
	(e\E \o)^2\, \mbfH^{(3)}\cdot \mbfH^{(1)} \sum_{n=0}^\infty  \a^{\pm,{(1)}} (t e\E\mbfH^{(1)} \cdot t e\E\mbfH^{(1)})^n
\end{equation}
and 
\begin{equation}\label{Expansion_TimeDependent_NLO2}
	(e\E \o)^2\, \mbfH^{(2)}\cdot\mbfH^{(2)} \sum_{n=0}^\infty  \a^{\pm,{(2)}} \left(te\E\mbfH^{(1)} \cdot t e\E\mbfH^{(1)}\right)^n
\end{equation}
Then, re-scaling $t\rightarrow \frac{t}{e\E|\mbfH^{(1)}|}$ as before, we express $O(\o^2)$ correction to the effective actions as
\begin{align}\label{EffectiveAction_Expansion_ElectricFake_NLO}
	\G^\pm_{\mrmE,\o^2} = -i\,  \o^2 \mcalV_3 \int \mrmd x_0\, \int_{J^\pm_0}  \frac{\mrmd t}{4\pi^2}\,  e^{\mp\frac{im^2 t}{e\E |\mbfH^{(1)}|}} \left[ \frac{e\E \mbfH^{(3)}\cdot \mbfH^{(1)}}{|\mbfH^{(1)}|} \,  \mcalS^\pm_{\o^2,1}(t) + \frac{e\E |\mbfH^{(2)}|^2 }{ |\mbfH^{(1)}|} \, \mcalS_{\o^2,2}^\pm(t) \right]
\end{align}
where  
\begin{equation}\label{Expansions_Rescaled_time}
	\mcalS_{\o^2,1}^\pm(t) = \sum_{n=0}^\infty  \a_{2,n}^{\pm,(1)} t^{2n} \quad , \quad \mcalS_{\o^2,2}^\pm(t) = \sum_{n=0}^\infty  \a^{\pm,(2)}_{2,n} t^{2n}
\end{equation}
with $\a_{2,n}^{\pm,(1)}$ and $\a_{2,n}^{\pm,(2)}$ being rational numbers. Again we provide first few terms for $\a_{2,n}^{\pm,(1)}$:
\begin{equation}\label{ExpansionCoeff_time_NLO1}
	\pm \left\{ \frac{1}{120}, - \frac{13 }{20160},\frac{157 }{4838400}, \frac{577 }{425779200},- \frac{2844701}{55794106368000}, \frac{2401579 }{1339058552832000}\right\}
\end{equation}
and for $\a_{2,n}^{\pm,(2)}$:
\begin{equation}\label{ExpansionCoeff_time_NLO2}
	\pm \left\{\frac{1}{160},- \frac{103}{80640},  \frac{2039}{19353600}, - \frac{1163}{189235200}, \frac{66417431}{223176425472000}, -\frac{6241699}{486930382848000} \right\}.
\end{equation} 
Using the terms up to $O(t^{48})$, we approximated the singularity structure associated to both series and found the leading poles again at $t^\pm_\star = \pm 2 \pi i$ with $O(10^{-8})$ precision. Note that contrary to the leading order poles, these are not simple poles. Instead, the singularity for associated to $\mcalS^\pm_{\o^2,1}(t)$ is a $3^\rd$ order pole while the one for $\mcalS^\pm_{\o^2,2}(t)$ is a $4^\th$ order one. Then, together with the prefactors, we get the leading order behaviour of the residues as
\begin{equation}
	R^\pm_{\o^2,1} \simeq -i\,0.11936621 \, \frac{ m^4}{(e\E|\mbfH^{(1)}|)^2} e^{-\frac{2\pi  m^2}{e\E|\mbfH^{(1)}|}} \simeq - \frac{3 i m^4}{8\pi\,(e\E|\mbfH^{(1)}|)^2}\, e^{-\frac{2\pi i m^2}{e\E|\mbfH^{(1)}|}}
\end{equation}
for the series $\mcalS_{\o^2,1}^\pm(t)$ and
\begin{equation}
	R^\pm_{\o^2,2} \simeq \mp 0.50000000  \, \frac{ m^6 }{(e\E |\mbfH^{(1)}|)^3} \, e^{-\frac{2\pi m^2}{e\E|\mbfH^{(1)}|}} \simeq \mp \frac{ m^6}{2(e\E|\mbfH^{(1)}|)^3}\, e^{-\frac{2\pi m^2}{e\E|\mbfH^{(1)}|}}
\end{equation}
for the series $\mcalS_{\o^2,2}^\pm$. Finally, considering the proper analytical continuation directions, we obtained the discontinuity for the effective action at order $O(\o^2)$ up $x_0$ integral as
\begin{equation}\label{LC_integral_time_NLO}
	\Im\D\G_{\mrmE,\o^2} \simeq \mcalV_3 \frac{ m^6\o^2}{4\pi (e\E)^2}\int \mrmd x_0\, \frac{ \mbfH^{(2)} \cdot \mbfH^{(2)}}{|\mbfH^{(1)}|^4}\, e^{-\frac{2\pi m^2}{e\E|\mbfH^{(1)}|}} 
\end{equation}
which only arises from the pole associated to the series $S^\pm_{\o^2,2}$.

As in the computation of the leading order term, we are going to handle the $x_0$ integral via the saddle point approximation based on the assumptions $m\gg e\E \mbfH^{(1)}(\o x_0)$ and the behaviour of $\mbfH^{(1)}(\o x_0)$ around its saddle points given in \eqref{Assumptions_time}. The first order saddle point approximation to \eqref{LC_integral_time_NLO} vanishes since $\mbfH^{(2)}(\o x_0^i) = 0$ for any saddle point $x_0^i$ and first non-zero term comes from the second order in the saddle point approximation. Although the second order approximation is more complicated than the first order (see e.g. \cite{bender1999advanced}), due to our assumption in \eqref{Assumptions_time}, only one term contributes to $\Im\D\G^{(2)}_{\mrmE,\o^2}$ and we found
\begin{equation}\label{SPA_NLO2}
	\Im\D\G^{(2)}_{\mrmE,\o^2} \simeq \mcalV_3\sum_i \frac{\o m^3}{8\pi^2\sqrt{e\E}} \frac{| \mbfh_3^{\{i\}}|^2\, |\mbfh_1^{\{i\}}|^{1/2}}{\left(\mbfh_3^{\{i\}} \cdot \mbfh_1^{\{i\}}\right)^{3/2}} \,  e^{-\frac{2\pi m^2}{e\E|\mbfh_0^i|}} .
\end{equation}
Note that contrary to singular behaviour of the leading order imaginary contribution in \eqref{PP_time}, the expression in \eqref{SPA_NLO2} vanishes in the uniform case. 

\subsection{Space Dependent Electric Fields:}
Now, we will consider space-dependent pure electric backgrounds for
\begin{equation}\label{GaugePotential_SpaceDependent}
	A_0(\mbfx) = \frac{\E}{\o}H_0(\o \mbfx).
\end{equation}
Since the effective dimensionality of this case depends on the details of the function $H_0$, its treatment with WKB methods is not always tractable. Therefore, the (proper) time dependent method we present here is much more convenient for applications. 

\textit{Notation:} As in the time dependent case, the number of derivatives acting on $A_0(\mbfx)$ counts the order in $\o$ expansion. However, now, due to the multi-dimensional nature of $A_0(\mbfx)$, a tensorial notation will be convenient when we tackle the saddle point approximation. For this reason we introduce the following notation:
\begin{equation}
	\frac{\dee^n \nabla A_0(\mbfx)}{\dee x_{k_1}\dee x_{k_2}\dots \dee x_{k_n}} = \mcalE \o^{n} \dee_{k_1}\dee_{k_2}\dots\dee_{k_n} \nabla H_0(\o \mbfx)
\end{equation}
where $\dee_{k_1}\dee_{k_2}\dots\dee_{k_n} \nabla H_0(\o \mbfx)$ is $n^\th$ order directional derivatives of $\nabla H_0$ evaluated at $\mbfx$. Note that we keep the first order derivative in vector notation rather than tensor notation since only $|\nabla H_0|$ appear in the exponent of the spatial integrals. Similar to the time-dependent case, we assume the following properties for the directional derivatives of $\nabla H_0$
\begin{align}
	i) \, \nabla H_0(\o \mbfx^i) &= \mbff^{\{i\}} \quad , \quad ii) \,  \dee_{k_1}\nabla H_0(\o \mbfx^i) = 0  \cr
	iii)& \, \dee_{k_1}\dee_{k_2}\nabla H_0(\o \mbfx^i) =  \mbff_{k_1,k_2}^{\{i\}} 
	. \label{Assumptions_spatial}
\end{align}
where $\mbff^{\{i\}}$ is a scalar corresponding to the value of $\nabla H_0$ at a saddle point $\mbfx^i$ and $\mbff_{k_1,k_2}$ is a vector corresponding to an element of the matrix\footnote{To be clear, the $2^{\nd}$ derivatives of $\nabla H_0(\o\mbfx)$ actually form a third rank tensor but since we will only need the vector valued function $f^{\{i\}}$ in future calculations, we kept the matrix form.} of $\nabla H_0$ evaluated at the same point.
\paragraph{Leading Order:} At the leading order, using the recursive relations \eqref{RecursionRelation_1}, we get the coupling expansion at order $O(\o^0)$ as 
\begin{equation}\label{Expansion_SpaceDependent_LO}
	\sum_{n=0}^{\infty} \a_{0,n}^\pm (t e\E \nabla H_0 \cdot te\E \nabla H_0)^n
\end{equation}
where the coefficients are the same with the ones in \eqref{ExpansionCoefficients_ElectricFake_1D}. 

Then, as in the time dependent case, rescaling $t\rightarrow \frac{t}{e\E |\nabla H_0|}$, we express the effective actions $\G^\pm_{\mrmE,\o^0}$ as
\begin{equation}
	\G^\pm_{\mrmE,\o^0} = - \mcalV_{\tD}\frac{(e \E |\nabla H_0|)^2}{4\pi^2} \int_{\mbbR^D} \mrmd^D x\, \int_{J^\pm_0} \frac{\mrmd t}{t^3} \, e^{\mp\frac{im^2 t}{e \E |\nabla H_0|} } \, \mcalS_{\o^0}^\pm(t).
\end{equation}
where $D=1,2,3$ depending to the potential $H_0(\o\mbfx)$ and $\tD = 4-D$ is the dimension of transverse directions including the time direction. Since the expansion is the same with the time dependent case, approximating it via Pade method leads to the leading pole at $t_\star^\pm = \pm 2\pi i n$ with residue
\[R^\pm_{\o^0} \simeq \frac{1}{4\pi^2}.\] Then, taking $t$ integral with proper analytical continuations leads to the same expression with the time dependent case up to the spatial dependence
\begin{equation}\label{LC_integral_space}
	\Im \G^\pm_{\mrmE,\o^0} = \pm \mcalV_{\tD} \int_{\mbbR^D} \mrmd^D x \, \frac{\left(e\E |\nabla H_0|\right)^2}{16\pi^3} \, e^{- \frac{2m^2 \pi} {e\E |\nabla H_0|}} .
\end{equation}
Finally, we handle the $D$-dimensional space integral via the saddle point approximation which leads to
\begin{equation}\label{SPA_LO_spatial}
	\Im\D \G_{\mrmE,\o^0} \simeq \mcalV_{\tD} \sum_i \frac{(e\E |\mbff^{\{i\}}|)^2}{8 \pi^3 }\left(\frac{ e\E |\mbff^{\{i\}}|^3}{m^2 \o^2 } \right)^{D/2} \, e^{-\frac{2m^2 \pi}{e\E |\mbff^{\{i\}}|}} \left(\det \bmh^{\{i\}} \right)^{-1/2}
\end{equation}
where 
\begin{equation}\label{HessianSPA_spatial}
	\bmh^{\{i\}}_{k_1,k_2} = \mbff_{k_1,k_2}^{\{i\}} \cdot \mbff^{\{i\}}
\end{equation}
is the matrix elements of $|\nabla H_0|=\sqrt{\nabla H_0 \cdot \nabla H_0}$ which are evaluated at a saddle point $x=x^{\{i\}}$.

\paragraph{Next to Leading Order:} 
As in the time dependent case, we organized the expansions at order $O(\o^2)$ in two parts:
\begin{equation}\label{Expansion_SpaceDependent_NLO1}
	(e\E \o)^2\, \nabla^3 H_0 \cdot \nabla H_0 \sum_{n=0}^\infty i \a^{\mp,{(1)}} (t e\E\nabla H_0 \cdot t e\E\nabla H_0)^n
\end{equation}
and 
\begin{equation}\label{Expansion_SpaceDependent_NLO2}
	(e\E \o)^2\,\left(\nabla^2\,H_{0}\right)^2 \sum_{n=0}^\infty i \a^{\mp,{(2)}} \left(te\E \nabla H_0 \cdot  t e\E \nabla H_0\right)^n
\end{equation}
and again re-scaling $t\rightarrow \frac{t }{e\E |\nabla H_0|}$, we get
\begin{align}\label{EffectiveAction_Expansion_ElectricSpatial_NLO}
	\G^\pm_{\mrmE,\o^2} = -i \o^2 \mcalV_{\tD}\int \mrmd^D x\, \int_0^\infty  \frac{\mrmd t}{4\pi^2}\,  e^{\mp\frac{im^2 t}{e\E |\nabla H_0|}} \left[ \frac{e\E \nabla (\nabla^2 H_0) \cdot \nabla H_0}{|\nabla H_0|} \,  \mcalS^\mp_{\o^2,1}(t) + \frac{ e \E \nabla^2 H_0  }{ |\nabla H_0|} \, \mcalS_{\o^2,2}^\mp(t) \right]
\end{align}
where $\mcalS_{\o^2,1}$ and $\mcalS_{\o^2,2}$ are written in \eqref{Expansions_Rescaled_time} with the same coefficients in \eqref{ExpansionCoeff_time_NLO1} and \eqref{ExpansionCoeff_time_NLO2} respectively but \textit{opposite} signs. The latter leads to the sign difference in the correction to the leading order contribution with respect to the time dependent case.

 Pade approximation to \eqref{EffectiveAction_Expansion_ElectricSpatial_NLO} leads to a similar expression with \eqref{LC_integral_time_NLO}
\begin{equation}\label{LC_integral_spatial_NLO}
	\Im\D\G_{\mrmE,\o^2} \simeq -\mcalV_{\tD}  \frac{ m^6\o^2 }{4\pi(e\E)^2} \int \mrmd^D x\, \frac{ |\nabla^2 H_0|}{|\nabla H_0|^4}\, e^{-\frac{2\pi m^2}{e\E|\nabla H_0|}} .
\end{equation}
Finally, in view of the assumptions in \eqref{Assumptions_spatial}, we get the first non-vanishing term as
\begin{equation}\label{SPA_spatial_NLO2}
	\Im\D\G_{\mrmE,\o^2}^{(2)} \simeq \mcalV_{\tD}\frac{m^4 \o^2\, }{8\pi^2 e\E |\mbff^{\{i\}}|} \left(\frac{e\E|\mbff^{\{i\}}|^3}{m^2\o^2}\right)^{D/2} \frac{\bmF^{\{i\}}\cdot (\bmh^{\{i\}})^{-1}}{\left(\det \bmh^{\{i\}}\right)^{1/2}} \, e^{\frac{-2m^2 \pi}{e\E|\mbff^{\{i\}}|}} ,
\end{equation}
where $\h$ is the matrix defined in \eqref{HessianSPA_spatial} and  $\tilde{\bm{\F}}^{\{i\}}$ is a matrix whose elements are given as \[\bmF^{\{i\}} = \sum_l \mbff^{\{i\}}_{l k_1} \cdot \mbff^{\{i\}}_{l k_2}.\] 

\subsection{Comments and Comparisons:} Now combining our results in \eqref{PP_time} 
and \eqref{SPA_NLO2}, the non-perturbative pair-production probability, i.e. $\mcalP_\mrmE = \frac{\D\G_{\mrmE}}{\mcalV_{\tD}}$, for a general time dependent electric field is written as
\begin{align}
	\mcalP_{\mathrm{time}} &\simeq  \sum_i \frac{(e\E)^{5/2}}{8\pi^3 m\o }  \sqrt{\frac{|\mbfh_1^{\{i\}}|}{\mbfh_1^{\{i\}}\cdot \mbfh_3^{\{i\}}} }  \left[|\mbfh^i_1|^3 +  \frac{m^4 \o^2 \pi}{(e\E)^3} \frac{|\mbfh_3^{\{i\}}|^2 }{\mbfh_3^{\{i\}} \cdot\mbfh_1^{\{i\}}} \right] e^{-\frac{2m\pi}{e\E|\mbfh_1^i|}}\label{PP_time_2orders}.
\end{align}
Similarly for the space dependent case, from \eqref{SPA_LO_spatial} 
and \eqref{SPA_spatial_NLO2} we get,
\begin{align}
	\mcalP_\mathrm{space} &\simeq \sum_i \frac{(e\E)^2}{8\pi^3 |\mbff^{\{i\}}|} \left(\frac{e\E |\mbff^{\{i\}}|}{m^2 \o^2}\right)^{D/2}\, \frac{ |\mbff^{\{i\}}|^3 - \frac{m^4 \o^2 \pi}{(e\E)^3} \bmF^{\{i\}} \cdot (\bmh^{\{i\}})^{-1} }{\left(\det \bmh^{\{i\}}\right)^{-1/2}} \, e^{-\frac{2\pi m^2}{e\E|\mbff^{\{i\}}|}}  \label{PP_spatial_2orders}
\end{align}

As a consistency check, we consider the one dimensional limits of both cases. Let the background electric field to be one dimensional and is directed along the $x_1$-axis. Then, the gauge fields are simplified to \[\mbfA(x_0) = \frac{\E}{\o}\left(H_1(\o x_0),0,0\right) \] and
\[A_0(\mbfx) = \frac{\E}{\o} H_0(\o x_1). \] Then, setting $\mbfh_1^{\{i\}} = \mbff^{\{i\}} = h_1^{\{i\}}$ and $\mbfh_3^{\{i\}} = \mbff_{k_1 k_2}^{\{i\}} = h_1^{\{i\}}$ in the one dimensional limit \eqref{PP_time_2orders} and \eqref{PP_spatial_2orders} simplifies to 
\begin{align}
	\mcalP^\pm_{1\mrmD} \simeq \sum_i \frac{(e\E)^{5/2}}{8\pi^3 m\o\sqrt{h_3^{\{i\}}}} \left[(h_1^{\{i\}})^3 \pm \frac{m^4 \o^2 \pi}{2(e\E)^3}\frac{h_3^{\{i\}}}{h_1^{\{i\}}} \right]\, e^{-\frac{2\pi m^2}{e\E h_1^{\{i\}}}}, \label{PP_1D_2orders}
\end{align}
where $\mcalP^+_{1\mrmD}$ and $\mcalP_{1\mcalD}^-$ refers to time and space dependent cases respectively.
\paragraph{Comparison to Worldline Instantons:} We can compare \eqref{PP_1D_2orders} with the well-known worldline instanton calculations \cite{Dunne:2005sx,Dunne:2006st} for the periodic and single-pulse background potentials: For the periodic case, i.e. $A(x) = \frac{\E}{\o} \sin(\o x) $, the saddle point relevant to the pair production rate is at $x=0$ where $h_1 = 1$ and $h_3 = 1$. Then, we get
\begin{equation}\label{PP_periodic}
	\mcalP_{\mathrm{periodic}} \simeq  \frac{(e\E)^{5/2}}{8 \pi^3 m\o} \left(1 \pm \frac{m^4 \o^2 \pi}{2(eE)^3}\right) e^{-\frac{2\pi m^2}{e\E}}.
\end{equation}
For the single-pulse case, i.e. $A(x) = \frac{\E}{\o}\tanh(\o x)$, again the saddle point approximation is dominated by $x=0$ where $h_1=1$ and $h_3=2$ and we get
\begin{equation}\label{PP_pulse}
	\mcalP_{\mathrm{pulse}} \simeq \frac{(e\E)^{5/2}}{8 \pi^3 m\o\sqrt{2}} \left(1 \pm \frac{m^4 \o^2 \pi}{(eE)^3}\right) e^{-\frac{2\pi m^2}{e\E}}.
\end{equation}
 Let us compare these results from the WL instantons for both periodic and single-pulse cases which were first computed in \cite{Dunne:2005sx,Dunne:2006st} for Euclidean proper-time case. In our convention of the Hamiltonian \eqref{Hamiltoian}, we get the imaginary part of the effective action for the periodic background as 
\begin{equation}\label{PP_WL_periodic}
	\Im \G^{\mathrm{WL}}_{\mathrm{periodic}}\simeq \mcalV_3 \frac{\sqrt{2\pi}(e\E)^{3/2}}{32 \pi^2} \frac{\left(1 \pm 2\g_\mrmE^2\right)^{3/4}}{\mbbK\left(\frac{\pm2\g_\mrmE^2}{1\pm 2\g_\mrmE^2}\right)} \,\frac{\exp\left\{-\frac{e\E}{\sqrt{2}\o^2} S_\mathrm{I} \right\}}{\mbbK\left( \frac{\pm 2\g_\mrmE^2}{1\pm2\g_\mrmE^2} \right) - \mbbE \left( \frac{\pm 2\g_\mrmE^2}{1\pm 2\g_\mrmE^2} \right)}
\end{equation}
where 
\begin{align}
	S_\mathrm{I} &= 4\sqrt{2} \sqrt{1\pm 2\g_\mrmE^2} \left[ \mbbK\left( \frac{\pm 2\g_\mrmE^2}{1\pm 2\g_\mrmE^2} \right) - \mbbE \left( \frac{\pm 2\g_\mrmE^2}{1\pm2\g_\mrmE^2} \right) \right] \cr
	& = -4\sqrt{2} \bigg[\mbbE(\mp 2\g_\mrmE^2) - (1\pm 2\g_\mrmE^2)\mbbK(\mp 2\g_\mrmE^2) \bigg] \label{ActionWL_periodic}
\end{align}
and single-pulse background as
\begin{equation}\label{PP_WL_pulse}
	\Im \G_{\mathrm{pulse}}^{\mathrm{WL}} \simeq \mcalV_3 \frac{(e\E)^{5/2}}{8\pi^3 m \o}\left(1\pm 2\g_\mrmE^2\right)^{5/4}\exp\left\{ -\frac{m^2 \pi}{e\E}\left(\frac{2}{1 + \sqrt{1 \pm  2\g_\mrmE^2}}\right) \right\}.
\end{equation}
In both cases, $\g_\mrmE = \frac{m\o}{e\E}$ and it is called the Keldysh parameter. The expansions in \eqref{PP_periodic} and \eqref{PP_pulse} match \textit{exactly} with the first two order expansions of \eqref{PP_WL_periodic} and \eqref{PP_WL_pulse} respectively. 

Note that the importance of the sign difference between the spatial and temporal cases are apparent in \eqref{PP_WL_periodic} and \eqref{PP_WL_pulse}: Both equations vanishes in $2\g_\mrmE \rightarrow 1$ limit for the spatial case while for the temporal case, they are always positive for any values of $\g_\mrmE^2>0$. This features are related to (lack of) a critical field strength which sets a lower bound for the creation of real particles for (temporal) spatial inhomogeneities.

In this sense, $O(\o^2)$ terms in \eqref{PP_periodic} and \eqref{PP_pulse} are the first corrections in the derivative expansion of the \textit{leading order} fluctuations of the worldline instantons. While on the one side, there are corrections to the derivative expansion, on the other side, there are also higher order corrections to the fluctuations which forms a series in $\frac{e\E}{2\pi m^2}$. Then, as in the recursive perturbative expansion that we discussed in the beginning of this chapter, the perturbative expansion around the worldline instantons is also a double expansion and the expansions in \eqref{PP_periodic} and \eqref{PP_pulse} corresponds to the first two terms in $o \ll e\E$ limit. It seems it is possible to obtain more information about other limits of these double expansions. This requires handling the recursive expansion in different way and utilizing the Pade method with a conformal transformation. (See e.g \cite{Dunne:2022esi}.) We postpone this problem for future work.

\section{Connection to WKB Formalism}\label{Section: SC_Connection} 
At the end of the previous section, we discussed that the non-perturbative pair production probabilities we obtained from the leading singularity of the Schrodinger kernel $\Tr e^{\mp i H t}$ matches with the WLI computations in \cite{Dunne:2005sx,Dunne:2006st}. From this equivalence we also observe, as it was conjectured by Andr\'{e} Voros \cite{voros94}, a one to one correspondence between the location of the singularities and the values of WLI actions. Another semi-classical method frequently used in the pair production problems is the WKB formalism where the information about the particle creation is obtained from the WKB loop integrals which are shown to be equivalent to WLI action \cite{Kim:2019yts,Taya:2020dco}. In the following, we discuss the connection between the time dependent setting we have been using in this paper and WKB formalism. This also helps us to elaborate on the connection of our discussion to WLI.

\paragraph{\underline{Setup}:} In this section, for concreteness, we focus on a specific background potential and choose the periodic  magnetic and (temporal) electric background fields in $x_3$ direction, i.e. 
\begin{equation}\label{MagneticGauge}
	A^{\mrmB}_\m = \left(0 , 0, -\frac{\E}{\o}\sin(\o x_1)  , 0 \right)
\end{equation}
and
\begin{equation}\label{ElectricGauge}
	A^{\mrmE}_\m = \left(0, 0, 0, -\frac{\E}{\o}\sin(\o x_0)\right)
\end{equation}

In addition to this, we only keep $p_0$ and $ p_1$ for the magnetic case and $p_0$ for the electric case non-zero and set the other components of momenta to zero as they only play a role in the pre-factor. Then, the corresponding Hamiltonians are
\begin{align}
	H_\mathrm{B} & = -\frac{p_0^2}{2} + \frac{p_1^2}{2} + \frac{g_\mrmB^2}{2\o^2}\sin^2(\o x_1)\quad , \quad g_\mrmB = e \B \label{Alternating_Magnetic}\\
	H_\mathrm{E} &= -\frac{p_0^2}{2} + \frac{g_\mrmE^2}{2\o^2}\sin^2(\o x_0) \qquad \; \quad  , \quad g_\mrmE = e\E \label{Alternating_Electric}.
\end{align}
In the magnetic case, we also defined a new variable as $-k^2 = m^2 - \frac{p_0^2}{2} <0$. Then, at the leading order in the recursive expansions \eqref{QuantumAction_timeDependent} and \eqref{QuantumAction_spaceDependent}, the action for the effectively one dimensional systems are written as 
\begin{align}
	\G^\pm_{\mrmB,0}(k^2) &= \mp  \int^{-k^2} \mrmd z \, \int_{J^\pm_0} \frac{\mrmd t\, e^{\mp i z t}}{2\pi}\, \int_{C^\pm_\mrmB} \mrmd x \, e^{\mp i t \frac{e^2 \mcalB^2}{2\o^2} \sin^2 (\o x)} \int_{C_p^\pm}\mrmd p  \, e^{\mp \frac{i t p^2}{2}}, \label{QuantumAction_magnetic}\\
	\G^\pm_{\mrmE,0}(m^2) & = \mp \int^{m^2} \mrmd z \, \int_{J^\pm_0} \frac{\mrmd t\, e^{\mp i z t}}{2\pi} \int_{C^\pm_\mrmE}  \mrmd x \, e^{\mp i t \frac{e^2 \mcalE^2}{2\o^2}\sin^2 (\o x)} \int_{C_p^\mp}\mrmd p  \, e^{\pm \frac{i t p^2}{2}}, \label{QuantumAction_electric}
\end{align}
where we dropped the space-time indices for simplicity. 

In both \eqref{QuantumAction_magnetic} and \eqref{QuantumAction_electric}, the $t$ variable is analytically continued in $\Im t<0$ and $\Im t>0$ for $\G^+$ and $\G^-$. This affects the region where each expression is well-defined. For example the $t$ integral for $\G_{\mrmB,0}^\pm$ is well-defined if 
\begin{equation}
	\Re\left[-k^2 +\frac{p^2}{2} + \frac{e^2 \B^2}{2\o^2}\sin^2(\o x)\right] \geq 0 , \label{MagneticCondition}
\end{equation}
while in the electric case, the condition becomes 
\begin{equation}
	\Re\left[m^2 - \frac{p^2}{2} + \frac{e^2\E^2}{2\o^2} \sin^2(\o x)\right] \geq 0 . \label{ElectricCondition}
\end{equation}
Note that in the limits when the left hand side of these expressions are zero, we encounter with a singularity of the classical limits of the Green's operators\footnote{Here, with the classical limit,  we mean that in the effective field theory perspective, there is no quantum correction to the classical potential $V(x)=\sin^2 x$ which changes the location of the singularities.} $G_{\mrmB,0} = (z + H_\mrmB)^{-1}$ and $G_{\mrmE,0} = \left(z + H_\mrmE\right)^{-1}$ respectively. From the classical mechanics point of view, $z$ can be viewed as energy variable. Then, in this sense, these singularities are associated to the (possibly complex) \textit{turning points} of the classical motion which is governed by the corresponding Hamiltonian. As we will see these points are the sources of the non-trivial information when we express $\G^\pm_{\mrmB,0}$ and $\G^\pm_{\mrmE,0}$ in terms of the WKB integrals.

Note that these singular limits also helps us to understanding the classical dynamics of each problem: Since both $k^2>0$ and $m^2>0$, for the magnetic case, we have the equality $k^2 = \frac{p^2}{2} + \frac{(e\B)^2}{2\o^2}\sin^2 x$ which is associated the motion of a particle inside a well of the inverse potential $V(x) = -\frac{(e\B)^2}{2\o^2}\sin^2 x$. On the other hand, for the electric case, we have $m^2 - \frac{p^2}{2} = \frac{(e\E)^2}{2\o^2}\sin^2 x$ which is linked to a scattering problem above the top of the potential barrier. These are compatible with the known properties of the magnetic and electric backgrounds. In the following, we will explain how these features are linked to the WKB formulation.

In order to make the connection with the WKB formalism clearer, we make same adjustments in our parameters. First, we introduce the Keldysh (adiabaticity) parameters as $\g_{\mrmB} = \frac{k\o}{e\B}$ and $\g_{\mrmE} = \frac{m\o}{e\E}$. Then, we re-scale \[t \rightarrow \frac{ 2\o^2}{g^2} t, \quad x \rightarrow \frac{x}{\o},\quad p \rightarrow \frac{g}{\sqrt{2}\o} p,\quad z \rightarrow \pm \frac{g^2 }{2\o^2} z,\] where $g = e\B$ for the magnetic case and $g=e\E$ for the electric case and $\pm$ signs in the last scaling are for electric and magnetic cases respectively.
\begin{align}
	\G^\pm_{\mrmB,0}(2\g^2_\mrmB) &= \pm \frac{e\mcalB}{\sqrt{2}\o^2} \int^{ 2\g^2_\mrmB} \mrmd z\, \int_{J^\mp_0} \frac{\mrmd t \, e^{\pm i z t}}{2\pi}\, \int_{C^\pm_\mrmB}\mrmd x\, e^{\mp it  \sin^2 x}  \int_{C_p^\mp}\mrmd p  \, e^{\mp \frac{i t p^2}{2}}, \label{QuantumAction_magneticScaled}\\
	\G^\pm_{\mrmE,0}(2\g^2_\mrmE) & = \mp \frac{e\mcalE}{\sqrt{2}\o^2} \int^{ 2\g^2_\mrmE} \mrmd z \,\int_{J^\pm_0} \frac{\mrmd t \, e^{\mp i z t}}{2\pi}\, \int_{C_\mrmE^\pm}  \mrmd x\,  e^{\mp it \sin^2 x}  \int_{C_p^\pm}\mrmd p  \, e^{\pm \frac{i t p^2}{2}}. \label{QuantumAction_electricScaled}
\end{align}

In the rest of this section,
\begin{itemize}
	\item We will first reduce $\eqref{QuantumAction_magneticScaled}$ and \eqref{QuantumAction_electricScaled} to WKB integrals. Then, we are going to associate each case to suitable WKB cycles and compute $\D\G_{\mrmB,0}(2\g_\mrmB^2)$ and $\D\G_{\mrmE,0}(2\g_\mrmE^2)$. 
	\item We will verify our results via reduction to the WKB formalism, we will also compute $\D\G_{\mrmE,0}$ and  $\D\G_{\mrmB,0}$, by directly taking the spatial integral along the corresponding Lefschetz thimbles. 
	\item Finally, we will finish with making a precise connection between our discussion on the electric case
	and EWKB approach.
\end{itemize}

\subsection{The WKB Integrals from the Propagator}\label{Section: Compare_WKB}
In this subsection, we will carefully map \eqref{QuantumAction_magneticScaled} and \eqref{QuantumAction_electricScaled} to WKB integrals by associating WKB cycles to electric and magnetic cases. For this reason, we will change the order of $t$ and $x$ integrals. This exchange is perfectly valid when the spatial contours $C_\mrmB^\pm$ and $C_\mrmE^\pm$ are chosen such that the integrals converges.

\begin{figure}[h]
	\centering
	\includegraphics[width=0.9\textwidth]{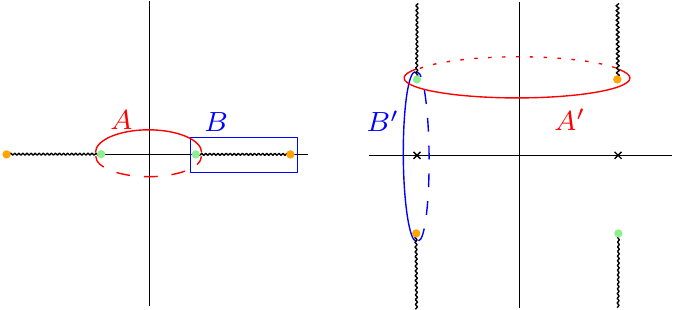}
	\caption{Possible WKB cycles for $V(x) = \sin^2 x$. \underline{Left}: Bounded region under the barrier top. \underline{Right}: Scattering region over the barrier top. In both cases, conventional WKB cycles are shown. They are named as {\color{red} $A$} and {\color{blue} $B$} cycles for the bounded region and {\color{red} $A'$} and {\color{blue} $B'$} cycles for the scattering region. Wavy lines are branch cuts. In the main text, we observe that a linear combination of {\color{red} $A'$} and {\color{blue} $B'$} cycles contribute to $\D\G_{\mrmE,0}$, while $\D\G_{\mrmB,0}$ is encoded by {\color{blue} $A$}-cycle.}	
	\label{Figure: WKB_cycles_periodic}
\end{figure}

Let us start with the electric case. In order to make the connection with WKB formalism clearer, we make the following manipulation \[z + \sin^2 x = \k - \cos^2 x =  \k  - \sin^2 x, \] where we defined $\k = z +1$ and shifted $x\rightarrow x + \frac{\pi}{2}$ to obtain the last expression. Then, the condition in \eqref{ElectricCondition} becomes
\begin{align}
	\Re\left[\k - \frac{p^2}{2}\right] \geq \Re\left[\sin^2 x\right]. \label{Condition_Electric}
\end{align}
In the case of $\k < 1$, \eqref{Condition_Electric} indicates the \textit{classically allowed} region of below the barrier top of potential $V(x)=\sin^2 x$. In this case, all the turning points  are real. In our original problem, on the other hand, we are interested in the scattering region i.e. $\k >1$, where the turning points become complex. We illustrated both cases in Fig. \ref{Figure: WKB_cycles_periodic}.  

In order to obtain the WKB integrals, we take the momentum\footnote{The proper way to take the momentum integrals is complexifying $p$ variable and rotating the initial contour to the directions where the integrand decays exponentially. This is a very special version of the Lefschetz thimble approach to the saddle point approximation. We will briefly discuss Lefschetz thimbles when we  deal with the spatial integral in Section \ref{Section: Thimbles}.} and time integrals in \eqref{QuantumAction_electricScaled}, we write $\G^\pm_{\mrmE,0}$ as
\begin{align}
	\G^\pm_{\mrmE,0}(2\g^2_\mrmE)
	& = \frac{e\mcalE}{\sqrt{2}\o^2} \int^{1+2\g^2_\mrmE} \mrmd \k\, G_{\mrmE,0}^\pm(\k), \label{SemiClassical_QuantumAction}
\end{align}
where
\begin{align}
	G^\pm_{\mrmE,0}(\k) & = \mp \int_{\tC_\mrmE^\pm}  \mrmd x\,  \left[\lim_{\t\rightarrow 0}\int_{0}^\infty  \frac{\mrmd t\, e^{\pm i \t}\, }{\sqrt{\mp 2\pi i t \, e^{\pm i \t}}}\,e^{- t \left[\pm i e^{\pm i \t} \left(\k - \sin^2 x\right)\right] }\right] \cr 
	&= \mp \frac{1}{\sqrt{2}}\int_{\tC_\mrmE^\pm} \frac{\mrmd x}{\sqrt{\k - \sin^2 x}} \label{TraceResolvent_magnetic}.
\end{align}
Note that $\tC_\mrmE$ is the contour for the shifted potential.

The integrand has a branch cut associated to the square root term. We choose the cuts to be in the regions where $\k< \sin^2 x$. Then, for $\k<1$, the integration contour $\tC_\mrmE$ is well defined between two neighbouring cuts and we place it to the interval $\tC_\mrmE^\pm = \left[-x_\mrmE,x_\mrmE \right]$, where $x_\mrmE = \sin^{-1}\sqrt{\k}$. (See Fig. \ref{Figure: WKB_cycles_periodic} (Left) where $\pm x_\mrmE$ are illustrated by green points.) In this sense, the discontinuity $\D G_{\mrmE,0} = G^\pm_{\mrmE,0} - G^-_{\mrmE,0}$ can be expressed by the integration along a WKB cycle in this region, which we call $A$-cycle. To see this connection, let us express $\D G_{\mrmE,0}$ as
\begin{align}
	\D G_{\mrmE,0} &= -\frac{2}{\sqrt{2}}\int_{-x_\mrmE}^{x_\mrmE} \frac{\mrmd x}{\sqrt{\k - \sin^2 x}} \cr 
	& = -\frac{1}{\sqrt{2}} \left[\int_{-x_\mrmE}^{x_\mrmE} \frac{\mrmd x}{\sqrt{\k - \sin^2 x}}  - \int_{x_\mrmE}^{-x_\mrmE} \frac{\mrmd x\; (-1)}{\sqrt{\k - \sin^2 x}} \right]
\end{align}
Then, re-writing the second part in the $2^\nd$ Riemann sheet, we absorb $(-1)$ term into $\sqrt{\k - \sin^2 x}$ and obtain
\begin{equation}
	\D G_{\mrmE,0}(\k)  = - \frac{1}{\sqrt{2}}\oint_{A} \frac{\mrmd x}{\sqrt{\k- \sin^2 x}}, \label{WKB_Period_magnetic}
\end{equation}
where $A$ is chosen in clockwise direction. In the WKB formalism, this quantity is called the \textit{``period''} associated to the underlying oscillatory classical motion along the $A$-cycle.

Finally, integrating over $\k$, we get the discontinuity $\D \G_{\mrmE,0} = \G_{\mrmE,0}^+ - \G_{\mrmE,0}^-$ as
\begin{equation}
	\D\G_{\mrmE,0} = -\frac{e\mcalE}{\sqrt{2}\o^2}\, \sqrt{2} \oint_{A} \mrmd x\, \sqrt{1+2\g^2_\mrmE -  \sin^2 x} \label{Gap_magnetic_WKB}
\end{equation}
where the loop integral along with the pre-factor $\sqrt{2}$ is indeed the leading order WKB integral for the semi-classical \textit{``action''} of the quantum mechanical system. (See e.g. \cite{Basar_2017} for the definitions that we use in this section for the period and action.)

For the magnetic case, upon momentum and proper-time integrals of \eqref{QuantumAction_electricScaled} in the same way with the electric case, we get
\begin{align}
	\G_{\mrmB,0}^\pm (2\g^2_\mrmB) 
	&= \frac{e \mcalB}{\sqrt{2}\o^2} \int^{2\g^2_\mrmB} \mrmd z \, G_{\mrmB,0}^\pm(z),
\end{align}
where 
\begin{align}
	G_{\mrmB,0}^\pm &= \pm \lim_{\ve \rightarrow 0}\int_{C_\mrmB^\pm} \mrmd x\, \int^\infty_0 \frac{\mrmd t}{\sqrt{\pm 2 \pi i t}} \, e^{-t \left[\mp i \left( z - \sin^2 x\right) \right] } \cr 
	&=  \pm \frac{1}{\sqrt{2}}\int_{C_\mrmB^\pm} \frac{\mrmd x}{\sqrt{z - \sin^2 x}}. \label{TraceResolvent_electric}
\end{align}
This is the same expression with $G_{\mrmE,0}^\pm$ up to the sign difference. Therefore, $C_\mrmB^\pm$ lies in the classically allowed region of the inverted periodic potential and the WKB dynamics is encoded by the corresponding $A$-cycle.

With the above discussion, we reduced the phase space integrals \eqref{QuantumAction_magneticScaled} and \eqref{QuantumAction_electricScaled} to WKB integrals and associated each case with the WKB cycles in the classical allowed region. However, for the electric case this corresponds to $m^2<0$; therefore, it is an unphysical case. It is possible to obtain the WKB actions for $\k>1$ by analytical continuation from the results of $\k<1$ region. To see the connection to the WKB cycles for $m^2>0 \,(\k>1)$, on the other hand, we recall that $\D \G_{\mrmE,0}$ is associated to $A$-cycle which consists of a path going from $-x_\mrmE$ to $x_\mrmE$ in the $1^\st$ Riemann sheet and coming back in the $2^\nd$ sheet. In $\k>1$, however, the branch cuts are placed between complex turning points and complex infinities. Therefore, any WKB cycles connecting neighbouring turning points can contribute to $\D \G_{\mrmE,0}$. This suggests that a linear combination of $A'$ and $B'$ cycles in Fig. \ref{Figure: WKB_cycles_periodic} should contribute to $\D\G_{\mrmE,0}$. In this section, we will compute this contribution through the analytical continuation from the unphysical region. In Section \ref{Section: ConnectionEWKB}, using the EWKB approach, we will associated $\D\G_{\mrmE,0}$ to the WKB cycles in scattering region which also yields a precise relationship between WKB cycles and WLI for the specific problem we are discussing. 

Now finally, we compute $\D\G_{\mrmE,0}$ and $\D\G_{\mrmB,0}$. We start from the period integrals in \eqref{WKB_Period_magnetic} and \eqref{TraceResolvent_electric} using the relation \cite{ramanujan_part3}
\begin{equation}
	\int_0^{\sin^{-1}\sqrt{u}}  \frac{\mrmd x}{\sqrt{u - \sin^2 x}} = \frac{\pi}{2} \, {}_2F_1 \Big(\frac{1}{2}, \frac{1}{2} ; 1, u\Big)  
\end{equation}
we get
\begin{align}
	\D\G_{\mrmB,0}(2\g_\mrmB^2) & = \left(\frac{e\mcalE}{\sqrt{2}\o^2}\right)2\sqrt{2} \int^{2\g_\mrmB^2} \mrmd z\, {}_2F_1 \Big(\frac{1}{2}, \frac{1}{2} ; 1, z\Big) \label{ActionWKB_magnetic} \\
	\D\G_{\mrmE,0}(2\g_\mrmE^2) &= -\left(\frac{e\mcalE}{\sqrt{2}\o^2}\right)2\sqrt{2} \int^{1+2\g_\mrmE^2} \mrmd \k\, {}_2F_1 \Big(\frac{1}{2}, \frac{1}{2} ; 1, \k\Big) \label{ActionWKB_electric}
\end{align}
Hypergeometric function is analytic expect on the line $ [1,\infty)$. Therefore, it is straightforward to compute the integral \eqref{ActionWKB_magnetic}:
\begin{equation}\label{WKB_Magnetic_Action}
	\D\G_{\mrmB,0}(2\g_\mrmB^2) = \left(\frac{e\mcalE}{\sqrt{2}\o^2}\right)4\sqrt{2} \Big[\mbbE(2\g^2_\mrmE) - (1-2\g^2_\mrmE) \, \mbbK(2\g^2_\mrmE) \Big]
\end{equation}
which is a manifestly real function for $0<\g_\mrmE <1$. For the electric case, on the other hand, since we are interested in region $\k>1$, we need an analytical continuation \cite[Eqn. 15.2.3]{NIST:DLMF}: 
\begin{equation}\label{EllipticFunction_AnalyticalContinuation}
	 {}_2F_1 \Big(\frac{1}{2}, \frac{1}{2} ; 1, u\Big) \rightarrow  {}_2F_1 \Big(\frac{1}{2}, \frac{1}{2} ; 1, u\Big) - i  {}_2F_1 \Big(\frac{1}{2}, \frac{1}{2} ; 1, 1- u\Big)
\end{equation}
where we pick the sign which will be compatible with the definition \[\D G(\k) = \lim_{\ve\rightarrow 0}\left( G(\k - i\ve) - G(\k + i\ve)\right) .\]  Finally, taking $\k$ integral of the analytical continued expression, we get\footnote{Note that here, we defined the elliptic integral as $\mbbK(m) = \int_0^{\pi/2} \frac{\mrmd x}{\sqrt{1 - m \sin^2 x}}$ where $m$ is the elliptic parameter rather than the modulus $k^2 = m$. With this definition it is related to the hypergeometric function as $\mbbK(m) = \frac{\pi}{2} {}_2F_1 \Big(\frac{1}{2}, \frac{1}{2} ; 1, u\Big)$.}
\begin{equation}
	\D\G_{\mrmE,0}(2\g_\mrmE^2) = \left(\frac{e\mcalE}{\sqrt{2}\o^2}\right) \left[S_{\mathrm{R}} + S_{\mathrm{I}}\right], \label{WKB_Electric_Action}
\end{equation}
where 
\begin{align}
	S_{\mathrm{R}} &= -4\sqrt{2} \Big[\mbbE(1+2\g^2_\mrmE) - 2\g^2_\mrmE \mbbK(1+2\g^2_\mrmE) \Big], \label{ActionElectric_1} \\
	S_{\mathrm{I}} &= -i 4\sqrt{2} \Big[ \mbbE(-2\g^2_\mrmE) - \left( 1 + 2\g^2_\mrmE \right) \mbbK(-2\g^2_\mrmE) \Big] . \label{ActionElectric_2}
\end{align}
As we expected, upon an appropriate analytical continuation, $\D\G_{\mrmE,0}$ is expressed as a linear combination of two WKB cycles and its imaginary part
\begin{align}
	\Im \D\G_{\mrmE,0}(2\g^2_\mrmE) &= -\left(\frac{e\mcalE}{\sqrt{2}\o^2}\right) 4\sqrt{2} \Big[ \mbbE(-2\g^2_\mrmE) + \left( 1 + 2\g^2_\mrmE \right) \mbbK(-2\g^2_\mrmE) \Big] 
\end{align}
is equal to the exponent of the pair production probability given by the WLI instanton in \eqref{PP_WL_periodic}.

\subsection{Direct Computations via Lefschetz Thimbles}\label{Section: Thimbles}
In this subsection, we discuss another way to compute both $\D\G_{\mrmB,0}$ and $\D\G_{\mrmE,0}$ by taking the integrals in \eqref{QuantumAction_magneticScaled} and \eqref{QuantumAction_electricScaled} without changing their order and we evaluate the spatial integral along the associated Lefschetz thimbles. On the one hand, we will obtain the same $\D\G_{\mrmE,0}$ in \eqref{WKB_Electric_Action} and $\D\G_{\mrmB,0}$ in \eqref{WKB_Magnetic_Action} without a direct reference to WKB formulation. Therefore, we can see that the discussion in this subsection provide a Lefschetz thimble perspective to the WKB method. In addition to that as we will see in the following, in this approach the analytical continuation we imposed for the electric case in the previous subsection appears naturally. Thus, in the following, no manipulation for $z$-parameter will be needed.



Let us start with the computations of the spatial integrals. In both magnetic and electric cases, the space integrals are in the same form: 
\begin{equation}\label{spaceIntegral}
	I_x^\pm = \int \mrmd x\, e^{\mp i t \sin^2 x} \quad , \quad \arg t = 0.
\end{equation}
While, the exponent in \eqref{spaceIntegral} does not contain any small/large parameter, it is always possible to re-write the exponent as $\frac{i t}{\g^2_\mrmE}\sin^2(\o x)$. Therefore, in $\g_\mrmE\ll 1$ limit, it is reasonable to use the saddle point approximation for \eqref{spaceIntegral}. The proper way to compute the saddle point approximation is allowing $x$ to be a complex variable and choosing \textit{``good paths''}  appropriately such that the integrand decays exponentially at infinities on those paths, i.e. $e^{-t \sin^2 x}$ as $|x| \rightarrow \infty $. 
These contours are called the Lefschetz thimbles and in the following we will label them with $\mcalJ$. There is also another set of contours which diverge exponentially at infinities. We would call them anti-thimbles and label with $\mcalK$. 
Note that both thimbles and anti-thimbles are the most dominant paths in the saddle point approximation.

An important point is that the set of thimbles form a (relative homology) basis such that all other possible well-behaved paths can be written as a linear combination of them.  (See \cite{Witten:2010cx,Witten:2010zr} for a precise construction, \cite{Aniceto:2018bis} for a review and \cite{Cherman:2014ofa} for a simple example very similar to our case.) In this way, it is possible to reconstruct any reasonable path using the set of thimbles. 

Since \eqref{spaceIntegral} is only a one-dimensional integral, the saddle point analysis simply leads to the stationary phase contours which are given by\footnote{Note that in a generic $D$ dimensional setting, the line \eqref{stationaryPaths} would be generalized to a $D$ dimensional surface corresponding to a solution of a flow equation \cite{Witten:2010cx,Witten:2010zr,Aniceto:2018bis}. However, we would not need it in this example.}
\begin{equation}\label{stationaryPaths}
	\Re \sin^2 x = \mathrm{const}.
\end{equation}
The good paths, on the other hand, determined by the imaginary part of $\sin^2 x$ and they should behave differently in $I_x^+$ and $I_x^-$ as the real time limit is taken from different directions. As $|x|\rightarrow \infty$, the thimbles $\mcalJ^+$ obeys
\begin{equation}
	\Im [t \sin^2 x] < 0,
\end{equation}
while $\mcalJ^-$ obeys
\begin{equation}
	\Im [t \sin^2 x] > 0.
\end{equation}
The anti-thimbles $\mcalK^\pm$ obey the opposite conditions to $\mcalJ^\pm$ in their respective cases.  Therefore, the thimbles and anti-thimbles for $I_x^+$ and $I_x^-$ swap.
\begin{figure}
	\centering
	\begin{subfigure}[b]{0.45\textwidth}
		\includegraphics[width=\textwidth]{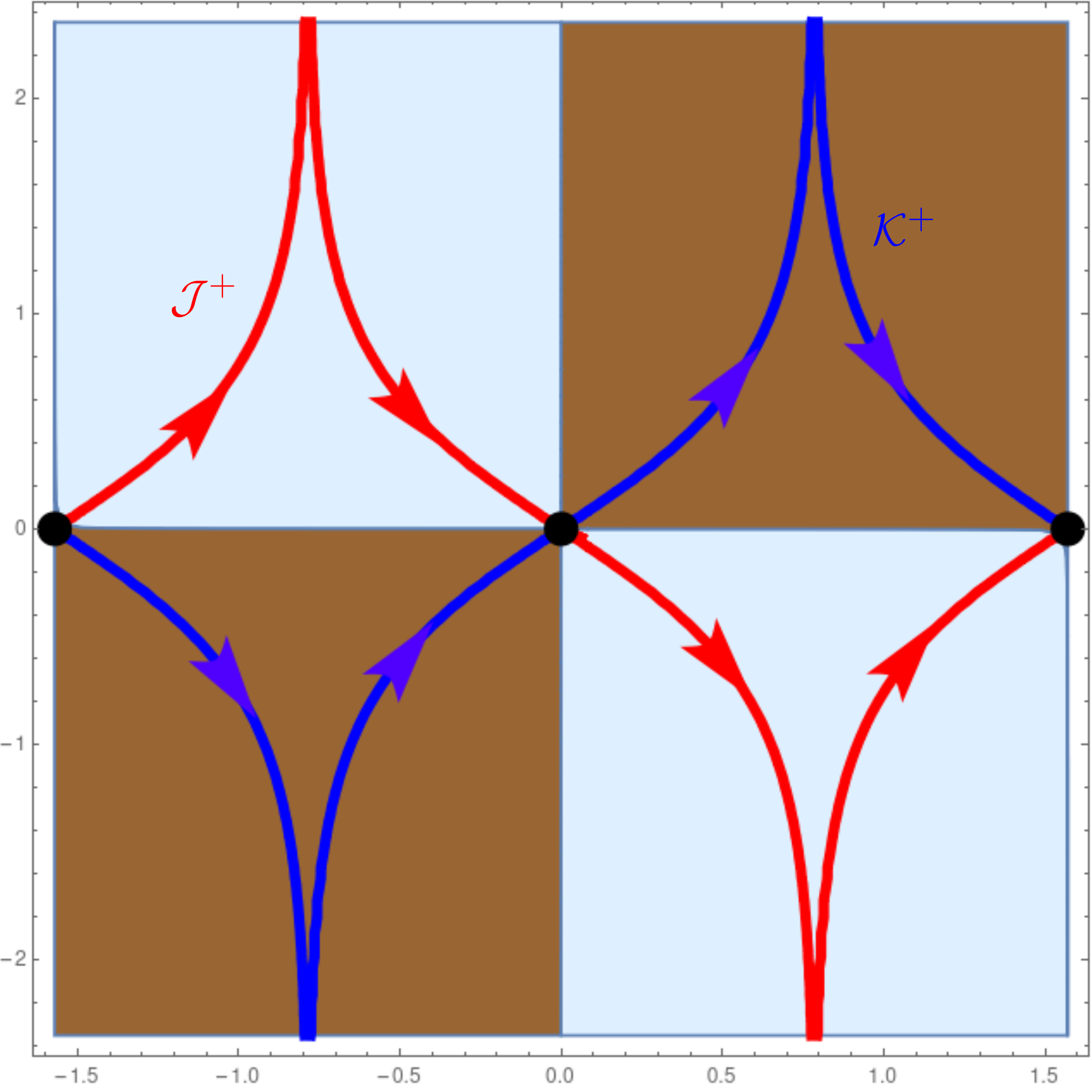}
		\caption{Thimble and anti-thimble for \textbf{forward} time evolution.}
		\label{Figure: ThimblesForward}
	\end{subfigure}
	\begin{subfigure}[b]{0.45\textwidth}
		\includegraphics[width=\textwidth]{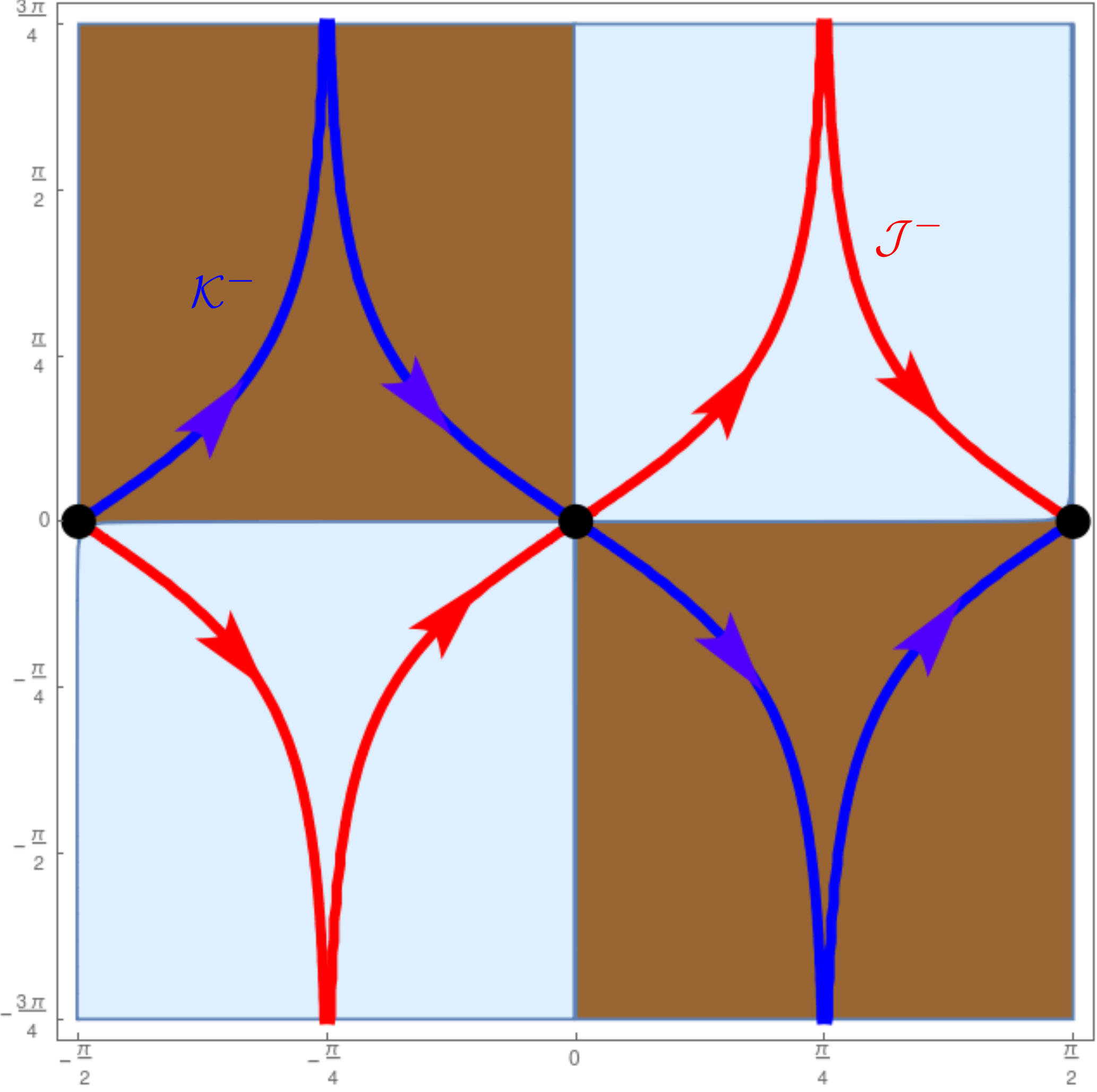}
		\caption{Thimble and anti-thimble for \textbf{backward} time evolution.}
		\label{Figure: ThimblesBackward}
	\end{subfigure}
	\caption{Lefschetz (anti-)thimbles for the spatial integrals of forward (left) and backward (right) time evolutions for both magnetic and the electric cases. In \textcolor{skyblue}{Lighter} regions the integrals converge while they diverge in \textcolor{darkbrown}{darker} regions. Paths for thimbles, i.e.  $\color{darkred}\mcalJ\pm$,  are plotted in \textcolor{darkred}{red} while paths for anti-thhimbles , i.e. $\color{darkblue}\mcalK^\pm$, are plotted in \textcolor{darkblue}{blue}. } 
	\label{Figure: LefschetzThimbles}
\end{figure}

Due to the periodic nature of the exponent, we can restrict ourselves to the region $\left[-\frac{\pi}{2},\frac{\pi}{2}\right]$. The thimbles for this region are given in Fig.~\ref{Figure: LefschetzThimbles}. Then, the effective path to the spatial integral can be obtain by a simple decomposition of $\mcalJ^\pm$ as
\begin{align}
	\mcalJ^\pm & = \left[-\frac{\pi}{2},\frac{\pi}{2}\right] + \left[-\frac{\pi}{4},-\frac{\pi}{4} \pm i\infty \right]  + \left[-\frac{\pi}{4} \pm i\infty , -\frac{\pi}{4}\right]\cr \cr
	& \quad + \left[ \frac{\pi}{4},\frac{\pi}{4} \mp i\infty\right] + \left[\frac{\pi}{4}\mp i\infty,\frac{\pi}{4}\right].
\end{align}
It is obvious that the lines in the imaginary direction cancel each other. Then, after the cancellations, we get
\begin{align}
	I_{x}^\pm(t) &= \int_{-\frac{\pi}{2}}^{\frac{\pi}{2}} \mrmd x\, e^{\mp i t\sin^2 x} \cr 
	& = e^{\mp\frac{i t}{2}} \int_0^\pi \mrmd x\, e^{\mp \frac{it}{2}\cos x} =  \pi e^{\mp \frac{i t}{2}} J_0\left(\frac{t}{2}\right) \label{spaceIntegral_magnetic}
\end{align} 
and re-write the effective action in \eqref{QuantumAction_electricScaled} as
\begin{align}
	\G_{\mrmE,0}^\pm (2\g_\mrmE^2)
	& = \frac{e\E}{\sqrt{2}\o^2}\int^{2\g_\mrmE^2} \mrmd z\, \left[\mp \sqrt{\frac{\pi}{2}} \int_0^\infty \frac{\mrmd t\, e^{\mp i z t}}{\sqrt{\mp i  t} } {}_1F_1\left(\frac{1}{2},1;\mp i t\right)\right] ,\label{ThimbleAction_Electric}
\end{align}
where ${}_1F_1\left(a,b;y\right)$ is the confluent hypergeometric function and we used \cite[Eqn. 10.16.5]{NIST:DLMF}
\begin{equation}
	e^{\mp \frac{i t}{2}} J_0\left(\frac{t}{2}\right) = {}_1F_1\left(\frac{1}{2},1;\mp i t\right)
\end{equation}
to obtain the final form of \eqref{ThimbleAction_Electric}. Note that the part in the square brackets is the trace of resolvent $G_{\mrmE,0}^\pm(z)$ in \eqref{TraceResolvent_magnetic}. Taking the time integral \cite[Eqn. 13.10.3]{NIST:DLMF}, we get
\begin{align}
	\G_{\mrmE,0}^\pm (2\g_\mrmE^2) &= \mp \left(\frac{e\mcalE}{\sqrt{2}\o^2}\right) \frac{\pi}{\sqrt{2}} \int^{2\g_\mrmE^2} \frac{\mrmd z}{\sqrt{z}} \, {}_2F_1\left(\frac{1}{2},\frac{1}{2},1;-\frac{1}{z}\right) \label{thimbleAction_integral0} 
\end{align}
Although, the integral can be directly taken since ${}_2F_1\left(\frac{1}{2},\frac{1}{2},1;-\frac{1}{z}\right)$ is analytic for $z>0$, it is instructive to re-write it as \cite{bateman1953higher,Fettis1970} 
\begin{equation}
	\G_{\mrmE,0}^\pm (2\g_\mrmE^2) = \mp \left(\frac{e\mcalE}{\sqrt{2}\o^2}\right) \frac{\pi}{\sqrt{2}} \int^{2\g_\mrmE^2}  \mrmd z\, \left[ {}_2F_1\left(\frac{1}{2},\frac{1}{2},1;1+z\right) \pm i \, {}_2F_1\left(\frac{1}{2},\frac{1}{2},1;-z\right)  \right]\label{thimbleAction_integral2}
\end{equation}
In \eqref{thimbleAction_integral2}, the term in the bracket together with $\mp \frac{\pi}{\sqrt{2}}$ term is $G^\pm_{\mrmE,0}$. Using the behaviour of ${}_2F_1\left(\frac{1}{2},\frac{1}{2},1;1+z \right)$ on its branch cut  \cite[Eqn. 15.2.3]{NIST:DLMF}, we observe $G_{\mrmE,0}^+=-G_{\mrmE,0}^-$. Then, we write $\D\G_{\mrmE,0}$ as
\begin{align}
	\D\G_{\mrmE,0} 
	& = - \left(\frac{e\mcalE}{\sqrt{2}\o^2}\right) \sqrt{2}\pi \int^{2\g^2_\mrmE} \mrmd z\,  \left[{}_2F_1\left(\frac{1}{2},\frac{1}{2},1;1+z \right) - i {}_2F_1\left(\frac{1}{2},\frac{1}{2},1;-z \right)\right]. \label{thimbleAction_integral3}
\end{align}
In this way, we recovered \eqref{ActionWKB_electric} with the analytical continuation given in \eqref{EllipticFunction_AnalyticalContinuation}. Finally, integrating the expression in the bracket, we recover \eqref{WKB_Electric_Action}
\begin{equation}
	\D\G_{\mrmE,0}(2\g_\mrmE^2) = \left(\frac{e\mcalE}{\sqrt{2}\o^2}\right) \left[S_{\mathrm{R}} + S_{\mathrm{I}}\right],
\end{equation}
where $S_{\mathrm{R}}$ and $S_{\mathrm{I}}$ are given in \eqref{ActionElectric_1} and \eqref{ActionElectric_2}.

Before, finishing this section, let us also compute $\D\G_{\mrmB,0}$ in the same fashion. Since the only difference between the phase space part of \eqref{QuantumAction_magneticScaled} and \eqref{QuantumAction_electricScaled} is the sign in the exponent of the momentum integrand, it is straight forward to get
\begin{align}
	\G_{\mrmB,0}^\pm &=  \pm \frac{e\B}{\sqrt{2}\o^2}\int^{2\g_\mrmB^2} \mrmd z\, \sqrt{\frac{\pi}{2}} \int_0^\infty \frac{\mrmd t\, e^{\pm i z t}}{\sqrt{\pm i  t} } {}_1F_1\left(\frac{1}{2},1;\mp i t\right)\\
	& = \pm \left(\frac{e\mcalB}{\sqrt{2}\o^2}\right) \frac{\pi}{\sqrt{2}} \int^{2\g_\mrmB^2} \frac{\mrmd z}{\sqrt{z}} \,\, {}_2F_1\left(\frac{1}{2},\frac{1}{2},1;\frac{1}{z}\right).
\end{align}
Then, re-writing the integrand using
\begin{equation}
	\frac{1}{\sqrt{z}} {}_2F_1\left(\frac{1}{2},\frac{1}{2},1;\frac{1}{z}\right) = {}_2F_1\left(\frac{1}{2},\frac{1}{2},1;z\right) \mp i {}_2F_1\left(\frac{1}{2},\frac{1}{2},1;1-z\right)
\end{equation}
we get
\begin{align}
	\D\G_{\mrmB,0} &= \left(\frac{e\mcalB}{\sqrt{2}\o^2}\right) \frac{\pi}{\sqrt{2}} \int^{2\g_\mrmB^2} \mrmd z\, {}_2F_1\left(\frac{1}{2},\frac{1}{2},1;z\right) \cr 
	& = \left(\frac{e\mcalB}{\sqrt{2}\o^2}\right) 4\sqrt{2} \left[\mbbE(2\g_\mrmB^2) - (1- 2\g_\mrmB^2) \mbbK(2\g_\mrmB^2)\right].
\end{align}
and recover \eqref{WKB_Magnetic_Action}.

\subsection{Connection to Exact WKB}\label{Section: ConnectionEWKB}
So far in this section, we computed the WKB action for both electric and magnetic backgrounds in two different ways: By reducing the classical limit of the effective action to WKB integrals and by direct computation along Lefschetz thimbles. Our findings are in accordance with the WLI computations \cite{Dunne:2005sx,Dunne:2006st} for the electric field and the lack of particle creation for the magnetic background. Now, we discuss how these results fits in the EWKB approach, which uses the Klein-Gordon equation rather than 1-loop effective action in path integral description.


On the time dependent electric field side, the EWKB approach was recently discussed in \cite{Taya:2020dco} where the particle creation was associated to the \textit{connection matrices} between different Stokes regions. In the setting presented in \cite{Taya:2020dco}, the connection problem is equivalent to the Bogoliubov transformation between the mode functions in asymptotic regions which satisfy the Klein-Gordon equation. However, the approach we presented throughout this paper is based on 1-loop effective action. Therefore, instead of connecting asymptotic regions, we will focus on the connection of successive turning points which is associated to the leading order WLI as described in \cite{Taya:2020dco}

For our choice of the gauge potential in \eqref{ElectricGauge} and $\hbar=1$, the Klein-Gordon equation is written as
\begin{align}
	0 =\left[\dee_0^2 + \frac{1}{2}\left(p_3 - \frac{e\E}{\o} \sin(\o x_0)\right)^2 +\frac{ p_\perp^2}{2} + m^2 \right]\f  
	\; \Longrightarrow \; \left(-\hbar^2\dee_0^2 + \sin^2 x_0  \right) \f = \k \f, \label{KleinGordon_Reduced}
\end{align}
where $\hbar = \frac{\sqrt{2}\o^2}{e\E}$, $\k = 1 + 2\g_\mrmE^2 $ and shifted $x_0$ by $\frac{\pi}{2}$. We also set $\mbfp=0$ to simplify the problem as before. Then, $e\E \gg \o$ limit  becomes the semi-classical limit, i.e $\hbar \ll 1$ 
of the Schrödinger equation on the right which formulates a scattering problem for periodic potential since $\k = 1 + 2 \g_\mrmE^2 >1$. In the following, we will discuss EWKB formalism very briefly with a focus on its connection to the semi-classical picture we illustrated in this section. The detailed introductory discussions on EWKB method can be found in \cite{kawai2005algebraic,Sueishi:2020rug,Enomoto:2020xlf,Taya:2020dco}.

In the EWKB perspective, the (formal) solutions of \eqref{KleinGordon_Reduced} are given as
\begin{equation}\label{EWKB_FormalSolution}
	\f_\pm(x_0) = \frac{1}{\s_\mathrm{even}}\,\exp\left\{\pm \int_{\tau}^{x_0} \mrmd x_0' \, \s_{\mathrm{even}} \right\},
\end{equation}
where $\tau$ is the reference point which is chosen to one of the turning points during computations and $S_\mathrm{even}$ is an infinite series in even powers of $\hbar$. Its leading order term is 
\begin{equation}
	\s_0 = \sqrt{\sin^2 x_0 - \k}
\end{equation}
and the rest of the series can be found recursively. These solutions have discontinuities, which are called \textit{Stokes lines} and they are given by 
\begin{equation}
	\Im \int_\tau^{x_0} \mrmd x_0'\, \sqrt{ \sin^2 x - \k} = 0,
\end{equation}
where $\tau$ is chosen as one of the turning points. 

\begin{figure}[h]
	\centering
	\includegraphics[width=0.9\textwidth]{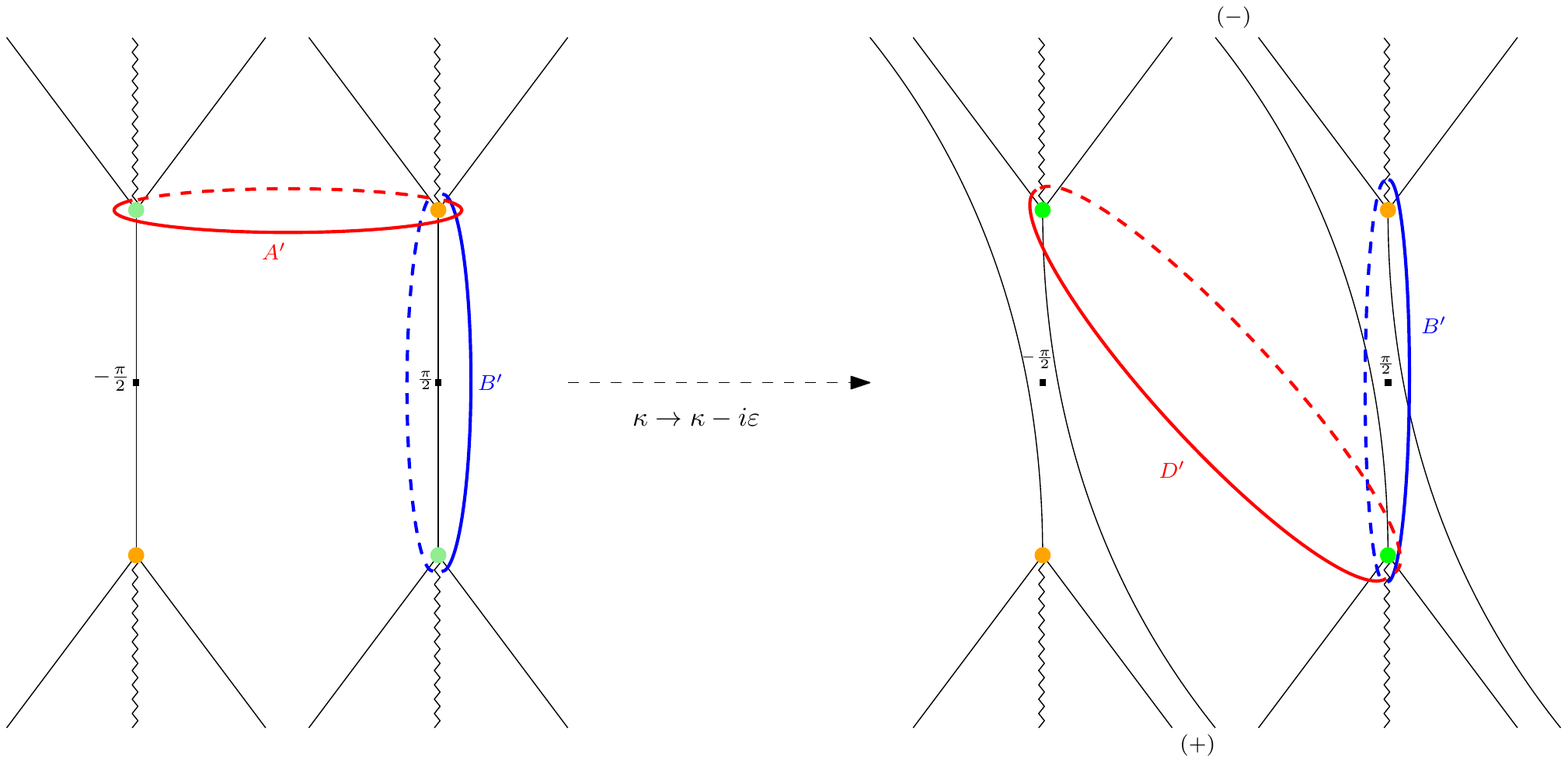}
	\caption{Stokes diagrams of $V(x) = \sin^2 x$ over the barrier region. Solid lines are Stokes lines emanating from the turning points and wavy lines are the branch cuts. \underline{Left}: Degenerate case. \underline{Right:} De-degenerate case with analytical continuation $\k \rightarrow \k - i\ve$ which is linked to the forward time evoluation. The cycles associated to the exact WKB procedure are denoted as {\color{red} D'} and {\color{blue} B'}. Note that this is different than the conventional picture in Fig. \ref{Figure: WKB_cycles_periodic} as ${\color{red} D'}={\color{red}A'}+{\color{blue} B'}$ and the information about the pair production process arises from ${\color{red} D'} - 2 {\color{blue} B'}$.
	}	
	\label{Figure: EWKB_StokesDiagrams}
\end{figure}

When $\k>1$, the turning points associated to the equation \eqref{KleinGordon_Reduced} in region $\left[-\frac{\pi}{2},\frac{\pi}{2}\right]$ are $x_\pm = \pm \frac{\pi}{2} + i \sin^{-1} \sqrt{\k}$ and their complex conjugates. The corresponding Stokes diagrams are illustrated in Fig. \ref{Figure: EWKB_StokesDiagrams}. Note that for real parameters, the Stokes lines are degenerate, i.e. a Stokes line originated from a turning point hits another turning point. In order the EWKB procedure works, these Stokes lines should be de-degenerated by an analytical continuation. 

Let us consider the choice $\k \rightarrow \k -i \ve$ which corresponds to the forward time evolution in the time dependent case. In this case, the connection of the solutions $\f_\pm\left(-\frac{\pi}{2}\right)$ to $\f_\pm\left(\frac{\pi}{2}\right)$ is given by the matrix
\begin{align}
	T^{(+)}&= \begin{pmatrix}	e^{\S_{x_+^* x_+}} & 0 \\ 0 & e^{-\S_{x_+^* x_+}}\end{pmatrix} \; 
	\begin{pmatrix} 1 \;\;& 0 \\ i \;\;& 1 \end{pmatrix} \;
	\begin{pmatrix}	e^{\S_{x_- x_+^*}} & 0 \\ 0 & e^{-\S_{x_- x_+^*}}\end{pmatrix} \\ \cr
	& = \begin{pmatrix}	
			e^{\S_{x_- x_+^*}+\S_{x_+^* x_+}} \;\;\;& 0 \\ 
			i\, e^{\S_{x_- x_+^*}-\S_{x_+^* x_+}}\;\;\; & e^{-\S_{x_- x_+^*}-\S_{x_+^* x_+}}
		\end{pmatrix}, \label{connectionMatrix1}
\end{align}
where $\S_{a b}$ is called the Voros coefficient and at the leading order semi-classical approximation it is given by
\begin{equation}
	\S_{a b} = \frac{1}{\hbar}\int_a^b \mrmd x \sqrt{\sin^2 x - \k}.
\end{equation} 

As explained in \cite{Taya:2020dco}, the pair production probability is related to $|T_{21}|^2$ whose exponent in \eqref{connectionMatrix1} can be written as
\begin{align}
	\S_{x_- x_+^*}-\S_{x_+^* x_+} &= + \frac{i}{\hbar} \left(\int_{x_-}^{x_+} \sqrt{\k - \sin^2 x} - 2 \int_{x_+^*}^{x_+} \sqrt{\k - \sin^2 x}\right) \label{exponentMatrixElement1}
\end{align}
where the choice of $(+)$ sign is compatible with $\k \rightarrow \k - i \ve$ choice: \[\sqrt{\sin^2 x - (\k - i\ve)} = \sqrt{e^{+ i \pi} \left(\k - i\ve - \sin^2 x \right) }= +i\sqrt{\k -i\ve - \sin^2 x}.\]
The integrals in \eqref{exponentMatrixElement1} are well-known \cite{Connor84}:
\begin{align}
	\int_{x_-}^{x_+} \sqrt{\k - \sin^2 x} &= 2\sqrt{2}\bigg[\Big(\mbbE(\k) - (1-\k)\mbbK(\k)\Big) - i \Big(\mbbE(1-\k) - \k \mbbK(1-\k) \Big) \Big] \label{Action_OverBarrierA}\\
	\int_{x_+^*}^{x_+} \sqrt{\k - \sin^2 x} &= -i 2\sqrt{2}\Big(\mbbE(1-\k) - \k \, \mbbK(1-\k)\Big), \label{Action_OverBarrierB}
\end{align}
where again the sign of the imaginary part in \eqref{Action_OverBarrierA} is chosen accordingly to $\k \rightarrow \k - i \ve$. Then, putting everything together and setting $\k = 1 + 2 \g_\mrmE^2$ and $\hbar = \frac{\sqrt{2}\o^2}{e\E}$, we get
\begin{equation}\label{cycles1st_sheet}
	\S_{x_- x_+^\star}-\S_{x_+^\star x_+}  = - \frac{i}{2} \left(S_\mrmR + S_\mrmI\right) = -\frac{i}{2}\D\G_{\mrmE,0}.
\end{equation}
where $S_\mrmR$ and $S_\mrmI$ are given in \eqref{ActionElectric_1} and \eqref{ActionElectric_2} respectively. Note that the right hand side of \eqref{cycles1st_sheet} is the contribution of half of the cycles $B'$ and $D'$ in the $1^\st$ Riemann sheet while $\D\G_{\mrmE,0}$ is the contribution of the whole cycles. With this given knowledge, we can get the leading order non-perturbative pair production probability as
\begin{equation}
	\mcalP = |T^{(+)}_{21}|^2 \sim e^{-\Im \D\G_{\mrmE,0}}, \label{PP_EWKB}
\end{equation}
which compatible with our previous discussion in Sections \ref{Section: Compare_WKB} and \ref{Section: Thimbles}. This also indicates that the contribution at 1 instanton level is equal to the sum of integration over $B'$ and $D'$ cycles in Fig. \ref{Figure: EWKB_StokesDiagrams} which can also be seen as a linear combination of $A'$ and $B'$ cycles. This is in complete agreement with the observation on WLI in \cite{Taya:2020dco}.

\section{Summary}

In this paper, we investigated the pair production problems using the real proper time approach. In the first part, we focused on keeping the final result unitary. Investigating electric fields with a general time and space dependence of the form $\mbfE = \E \frac{\mrmd H(\o x_0)}{\mrmd x_0}$ and $\mbfE = \E \nabla H_0(\o \mbfx)$, we computed the first two orders of the non-perturbative pair production probability in the locally constant field limit, i.e. $\o \ll 1$. Our method relies only on the perturbative expansion of the real proper time propagator $U^\pm = e^{\mp i H t}$ which we generated by a simple recursion relation. After summing this series via Pade approximation, we associated the pair production probability as a difference between the discontinuities of the effective actions of forward and backward time directions, i.e.  $\D\G_\mrmE = \G^+_{\mrmE}$ - $\G^-_{\mrmE}$. The discontinuities arises due to a Borel like integration over a singularity of $\Tr U^\pm(t)$ at a complex non-zero point in $t$-plane. In this sense, we showed that the singularity of $\Tr U^\pm$ acts as the source for the pair production process. 

In addition to computing the unitary preserved pair production probability, we also managed to understand how this real time approach is linked to the well known semi-classical approaches, namely the Euclidean worldline instantons and WKB methods. First, we found an exact agreement between our calculations and the Euclidean time worldline instanton calculations in \cite{Dunne:2005sx,Dunne:2006st} which is also a verification of a conjecture between action of the classical motion and the singularities of the propagator as conjectured by Andr\'e Voros \cite{voros94}.

Later, focusing on one dimensional periodic electric background potential, i.e. $\sin^2 x$, we showed how the WKB integrals appear in the classical limit of the real time dependent formulation and linked the WKB cycles of the potential $V(x)$ to the discontinuities of the electric and magnetic cases in $\hbar \rightarrow 0$ limit, which we denoted as  $\D\G_{\mrmE,0}$ and $\D\G_{\mrmB,0}$ respectively. In this way, we showed that the electric case is linked to a linear combination of WKB cycles in the scattering region of the potential while the magnetic case is represented by the perturbative WKB cycle inside the well of $\sin^2 x$. We computed both $\D\G_{\mrmE}$ and $\D\G_{\mrmB}$ by integrating the WKB integrals we obtained as well as taking the phase-space integrals directly along the Lefschetz thimbles without any direct reference to the WKB cycles.  For the electric case, $\Im \D\G_{\mrmE,0}$ matches with WLI action while we obtained real $\D\G_{\mrmB,0}$ which indicates the lack of particle creation as expected.  Note also that the thimble approach presents a new way to obtain the WKB actions which can be utilized in higher dimensional problems as well.

We completed our discussion with a connection to EWKB method and showed which specific WKB cycles are responsible with the particle creation. In complete harmony with the time dependent approach, we found that it is a linear combination of the two linearly independent cycles over the barrier region. Since the integration cycles in EWKB approach are linked to WLI \cite{Taya:2020dco}, this also provided a precise example on the connection between WLI and the classical action $\D\G_{\mrmE,0}$.

Note that the pair production probability $e^{-\Im \D\G_{\mrmE,0}}$ we get in EWKB approach offers an exponentiation of $\D\G_{\mrmE,0}$ by including higher order terms. It would be interesting to explore possible methods to reconstruct the exponential instanton contribution to the semi-classical limit of $\D\G_{\mrmE}$. Apart from very hard analytical explorations in this direction, it might also be desirable to rely on the perturbative expansions and making use of the conformal Pade methods \cite{Costin:2019xql, Costin:2020hwg, Costin:2020pcj, Florio:2019hzn} in an $\hbar$ expansion rather than a LCFA limit. On the one side, it would strengthen the connection between WKB formulation, WLI and singularity structure of $\Tr U^\pm$, it would enable to probe the singularity not only in $\g^2_\mrmE \gg 1$ limit but for all values of $\g_\mrmE^2$. (See \cite{Dunne:2022esi} for an example.)

\section*{Acknowledgment}
The author thanks Dieter Van den Bleeken, Can Kozçaz, İlmar Gahramanov and Mithat Ünsal for various discussions in the earlier stages of this research. He also thanks to Reiko Toriumi and Rudrajit Banerjee for discussions on the exact WKB method.

\appendix

\bibliographystyle{JHEP}
\bibliography{PP_Resurgence.bib}

\end{document}